\documentclass{aa}
\usepackage{graphicx}

\begin{document}

\title{The INES System IV: The IUE Absolute Flux Scale}

\author
{
R. Gonz\'alez-Riestra\inst{1}\fnmsep\thanks{\emph{Previously:}
ESA-IUE Observatory}
\and
A. Cassatella\inst{2,3}\fnmsep$^*$
\and
W. Wamsteker\inst{4}\fnmsep\thanks{Affiliated to the
Astrophysics Division, SSD, ESTEC}
}

\offprints
{
R. Gonz\'alez-Riestra,
\email{ch@laeff.esa.es}
}

\institute
{
Laboratorio de Astrof\'{\i}sica Espacial y F\'{\i}sica
Fundamental, VILSPA, P.O. Box 50727, 28080 Madrid, Spain
\and
Istituto di Astrofisica Spaziale, CNR, Area di Ricerca Tor
Vergata, Via del Fosso del Cavaliere 100, 00133 Roma, Italy
\and
Dipartimento di Fisica E. Amaldi, Universit\'a degli Studi Roma
Tre, Via della Vasca Navale 84, 00146 Roma, Italy
\and
ESA-IUE Observatory, VILSPA, P.O. Box 50727, 28080 Madrid, Spain
}

\date{Accepted: May 4, 2001}

\abstract 
{ 
This paper deals with the definition of the input fluxes used for the
calibration of the IUE Final Archive. The method adopted consists on the
determination of the {\it shape} of the detector's sensitivity curves using
IUE low resolution observations with model fluxes of the DA white dwarf
G191--B2B.  A {\it scale factor} was then determined so that the IUE
observations of some bright OAO--2 standards match the original
measurements from Meade (1978) in the spectral region 2100--2300 \AA. The
ultraviolet fluxes of six standard stars used as input for the Final
Archive photometric calibration together with the model fluxes of G191--B2B
normalized to the OAO--2 scale are given.  A comparison with the
independent FOS calibration, shows that the IUE flux scale for the
Ultraviolet is 7.2 \%\ lower. We consider this mainly to be caused by the
different normalization procedures. It is shown that the present flux
calibration applies to spectra processed with the {\it INES} low resolution
extraction software.
}

\authorrunning{R. Gonz\'alez-Riestra et al.}

\titlerunning{The IUE Flux scale}

\maketitle

\section{Introduction}
\label{sec:intro}

Many improvements have been made to the standard processing of IUE data
along the years. The combination of a better understanding of the
instruments and the rapid evolution of computing capabilities, has allowed
to use the carefully planned calibration data, obtained under well
controlled acquisition conditions over the 18 years of the IUE Project, to
prepare a new calibration of the complete IUE data set.

The IUE Final Archive, {\it IUEFA}, was the end-product of the above
process, which started to be defined in the late eighties. The processing
software developed for this purpose was {\it NEWSIPS} (Garhart et
al. 1997). The {\it INES} (IUE Newly Extracted Spectra) System is the final
configuration of the IUE archive.

A detailed revision of the {\it NEWSIPS} output products indicated that
there were still some problems which could be corrected.  The most
important deficiency was found in the {\it NEWSIPS} extraction and noise
models for low resolution spectra ({\it SWET}), which e.g. caused emission
line fluxes to be frequently wrongly extracted (Schartel and Skillen,
1998).  In high resolution data, a systematic mismatch of about 20
km~s$^{-1}$ between the velocity scales of short and long wavelength
spectra was present. These, together with other errors, were corrected in
the {\it INES} system developed by the ESA IUE Observatory (Wamsteker et
al. 2000). A full description of the {\it INES} system and its data
processing is given in Rodr\'{\i}guez-Pascual et al. (1999), Cassatella et
al. (2000) and Gonz\'alez-Riestra et al. (2000).  The {\it INES} Data are
available from the {\it INES} Principal Centre {\tt
http://ines.vilspa.esa.es} or from the {\it INES} National Hosts (Wamsteker
2000). For details on the instrumental history of IUE see P\'erez--Calpena
and Pepoy (1997).

In this paper we discuss the way the IUE absolute flux scale was redefined
(Sect.~\ref{sec:iueflux} and~ \ref{sec:fluxcal}). The specific algorithms
needed to optimize the internal consistency of IUE spectra, such as those
used to determine the effective exposure times and to correct for the time
and temperature dependency of the sensitivity of the IUE cameras are
described in Sec.~\ref{sec:fluxcal}. In Sec.~\ref{sec:hst}, a comparison is
made between fluxes obtained through the present calibration and those
derived from previous IUE calibrations and from other experiments (HST and
HUT). In Sec.~\ref{sec:ines} we demonstrate the applicability of the our
calibration to the data in the {\it INES} archive.

\section{The IUE Flux Scale}
\label{sec:iueflux}

Along the operational life of IUE, and prior to the Final Archive
processing, several photometric calibrations algorithms have been applied
as a consequence of the changes made in the processing software.  In all
cases the flux calibration was based on the UV absolute fluxes of the
bright B3~V standard star $\eta$ UMa (V=1.84) as defined by Bohlin et
al. (1980).  However, evidence for systematic errors in this $\eta$ UMa
flux scale made it necessary to find alternatives to be used as primary
calibration standards for the IUE Final Archive. In this Section we will
describe the basis of the early IUE photometric calibrations and the new
flux scale.

The primary flux calibration for IUE data is done on the low resolution
spectra, while the high resolution calibration is derived from this. The
common basis of all early calibrations was the absolute flux of $\eta$ UMa
defined by Bohlin et al.  (1980), who took the OAO-2 data as main reference
for fluxes longward 2000 \AA, and the rocket data of Brune et al. (1979)
for shorter wavelengths. $\eta$ UMa is too bright to be observed directly
with IUE at low dispersion, and therefore a set of secondary standard stars
was defined. These were chosen from the OAO--2 and TD1 Catalogues. The
original OAO--2 and TD1 fluxes of these standards were reduced to the
common $\eta$ UMa flux scale by applying the ``correction factors'' given
by Bohlin and Holm (1984).

With the growing observational material acquired over the years, it became
clear that there were systematic differences between observations and
models for objects of very different physical nature, such as white dwarfs
(Greenstein and Oke 1979), BL Lac objects and sdO stars (Hackney et
al. 1982). Finley et al. (1990) showed discrepancies of up to a 15\%\ when
comparing IUE observations and fluxes predicted by models of DA white
dwarfs.  The fact that these differences were maximum in the region of
largest disagreement between the original OAO--2 and TD1 fluxes, pointed to
the existence of systematic errors in the $\eta$ UMa flux scale.

A complete revision of the IUE flux calibration was therefore considered a
primary requirement in the planning of the IUE Final Archive (Cassatella
1990).  Rather than deriving the flux scale for the UV on a star which can
not be observed with the instrumental setup supplying the bulk of currently
available UV data, a different approach was taken, allowing to use the IUE
large data set and to make new special purpose observations to derive an
independent calibration.  Hot DA white dwarfs were chosen as the most
suitable objects to define the {\it relative} IUE flux scale. They were
used to determine the {\it shape} of sensitivity curves by comparison of
the IUE observations with model fluxes. A {\it scaling factor} was defined
to bring the relative fluxes of the OAO--2 standards at an absolute
scale. In the absence of other (and better) calibration sources for the
space--UV, the absolute scale was defined by the original OAO--2
measurements from Meade (1978). The accuracy on computed fluxes for DA
white dwarfs is discussed by Finley (1993).

To obtain the shape and the scale factor of the sensitivity curves, an
intensive observing campaign was made in 1990 and 1991. These observations
included not only the traditional TD1 and OAO--2 standards already in use,
but also a selected sample of DA white dwarfs.  Details of the procedure
followed to obtain the input fluxes for the IUE calibration are given in
the next Section.

\section{The calibration of the IUE Final Archive}
\label{sec:fluxcal}

\subsection{The input Data}

Two sets of data were used to derive the flux calibration for the IUE Final
Archive. The first one consisted of a large number of observations of the
IUE standard stars taken at the time of the acquisition of the 1984--85
Intensity Transfer Functions (hereafter ITFs). This set included spectra
obtained in all the possible observing modes (high and low dispersion,
large and small aperture, trailed, etc).

A considerably more extended set of calibration data was taken is 1991,
which included not only observations of the IUE standard stars, but also of
several selected white dwarfs, and in particular G191--B2B. The acquisition
of these data was carefully planned to determine all parameters necessary
for the calibration of the instruments, such as the size of the
spectrograph apertures and the camera response times.  The 1991 data were
used to derive the absolute fluxes of the IUE standard stars. The use of
close--in--time observations of both the white dwarfs and the standard
stars avoided the need to correct for the cameras sensitivity loss.

%========================================================================
\begin{table}
\begin{center}

\caption { }

\centerline{Number of spectra used to derive the}
\centerline{ Absolute Fluxes of the Standard Stars}

\begin{tabular}{l c c c c}
		&		&		&	    &	\\	
\hline
Wavelength 	&  G191-B2B	& Bright	& Faint	    &             \\
Range 		&		& Stars$^1$	& Stars$^2$ &	Total	\\
\hline
Short	&	19	& 39	& 45	& 103 \\
Long	&	19    	& 43	& 66	& 128 \\
\hline
	&		&		&	\\
\end{tabular}

\centerline{Number of spectra used to derive the}
\centerline{Inverse Sensitivity Curves of the IUE Cameras}

\begin{tabular}{l c c c}
	&		&		&	\\
\hline
	&	Bright 	  & Faint 	& 	\\
Camera	&	Stars$^1$ & Stars$^2$	& Total	\\
\hline
SWP 	&	29~~	&	104	&	135	\\
LWP 	&	14~~	&	91	&	105	\\
LWR 	&	22~~	&	41	&	63	\\
\hline
\end{tabular}
\label{tab:numb}
\end{center}
$^1$ $\eta$ Aur, $\lambda$ Lep, 10 Lac, $\zeta$ Dra \\
$^2$ BD+28~4211, BD+75~325, HD~60753 \\
\end{table}
%========================================================================

Only point--source Large Aperture spectra were used for the derivation of
the flux calibration. As part of the complete calibration of the IUE
instrument, the factors necessary to calibrate other observational modes,
e.g. trailed spectra, were redetermined.

Table \ref{tab:numb} gives the number of standard star spectra used to derive the
photometric calibration of the IUE cameras.

\subsection{The Intensity Transfer Functions (ITFs)}
\label{sec:itf}

The ITFs are used to linearize the IUE raw Data Numbers (DNs) by
transforming them into Flux Number (FNs). The ITFs are constructed from
graded exposures of lamps under well controlled thermal spacecraft
conditions and radiation background. For historical reasons these ITFs have
been derived through linear interpolation between 12 selected exposure
levels spaced over the dynamic range of the IUE Cameras (from 0 to 255
DN). This has made that some small linearity errors for the highest and
lowest exposure levels have persisted in the IUE data (Gonz\'alez--Riestra
1998).  Since the ITFs define the linearity of the cameras, any calibration
is linked to a specific ITF.
 
In what follows, we describe the ITFs used for the derivation of the IUE
Final Archive flux calibration.

\underline {LWP}: The original ITF for this camera was based on data
obtained in 1984--1985. It was decided to acquire a new ITF in May
1992. Although some anomalies were found in the cross--correlation behavior
of this ITF, its effects were limited and it was decided to maintain the
1992 ITF for the IUEFA processing. The existence of two well differentiated
groups of zero level (``NULL'') images, presented an additional anomaly.
Although the cross--correlation behavior of one of these two ( the
``NULL-A'') was worse, this was selected for the complete processing, since
it avoided strong negative extrapolations at the short wavelength end of
the camera.

\underline {LWR}: The LWR camera was declared non--primary long wavelength
camera in October 1983 (P\'erez--Calpena and Pepoy 1997).  A new ITF was
acquired one month later. It was found that this ITF gave a poor
correlation with science images, especially with those taken before the
camera was declared non--operational.  Two ITFs were  constructed for this
camera. ITF--A is the original 1983 ITF with its own NULL level. It is
appropriate for most of the images taken after 1983. ITF--B has as NULL
level the average of all the NULL images with similar geometric
characteristics taken in the period 1978--1983. The upper levels are the
same as in ITF--A, but resampled to match the geometric characteristics of
this modified NULL level. The {\it NEWSIPS} processing cross--correlates
every science image with both ITFs, choosing for the processing the one
with the highest correlation coefficient. The use of two different ITFs in
the photometric correction required to derive two inverse sensitivity
curves for this camera.

\underline {SWP}: The ITF acquired in 1985 was used for the processing of
all SWP images.

\subsection{The White Dwarf model}

The white dwarf G191--B2B was selected as primary standard to define the
relative fluxes of the other IUE standard stars due to its brightness
(V=11.8), pure hydrogen atmosphere, high effective temperature (implying a
narrow Lyman $\alpha$ absorption line), and negligible interstellar
absorption (N$_{\rm H}\approx$ 1.7~10$^{18}$ cm$^{-2}$, Kimble at
al. 1993).  The model used was provided by D. Finley (private
communication, 1991), and was computed using the code of D. Koester (see a
detailed description in Finley et al. 1997). The model has the following
characteristics (see Fig.~\ref{fig:model}):

\begin{itemize}

\item Chemical composition: Pure Hydrogen
\item T$_{\rm eff}$ = 58000 K
\item log~g = 7.5

\end{itemize}

Evidence for the presence of metals in this star has been reported by
Bruhweiler and Kondo (1981) and by Bruhweiler (1991) from IUE spectra, and
by Vennes (1992) and Barstow et al. (1993) from ROSAT
observations. However, the abundance of these elements is extremely low
(C/H=2$\times$10$^{-6}$, N/H=4$\times$10$^{-6}$, Si/H=1$\times$10$^{-6}$,
Fe/H=5$\times$10$^{-6}$, Ni/H=1$\times$10$^{-6}$; Wolff et al.  1998) and
their influence in the IUE range is negligible.  According to Finley
(1993), the overlapping metal lines might reduce the FUV continuum by
1--2\% in some spectral regions. The effective temperature and the gravity
were derived from the profiles of the optical Balmer lines (Finley, private
communication). The model provided by Finley was normalized to the
spectrophotometric data in the range 3200--8000 \AA\ as given by Massey et
al. (1988).

The particular choice made here for the model parameters of G191--B2B, has
a little effect on the IUE calibration in the sense that, if an improved
model becomes available in the future, it would be straightforward to
derive a suitable correction from the ratio between the new model and the
one used here (see Appendix B).

\subsection{Other parameters and algorithms}

\subsubsection{Determination of exposure times}
\label{sec:CRFT}

For very short exposures, the effective exposure time of the IUE cameras
(t$_{\rm eff}$) is different from the commanded one (t$_{\rm com}$), due to
the quantization of the clock (0.4096 sec.) and the so--called ``Camera
Rise/Fall time'' (CRFT). New data were obtained to re--derive the rise/fall
times in 1991 and the values used in the IUEFA production are given in
Table \ref{table:crt} (Gonz\'alez-Riestra 1991).  The effective exposure time
is:

\begin{equation}
\rm{t_{eff}=INT(t_{com}/0.4096)\times 0.4096 - t_{rise}}
\end{equation}

%+++++++++++++++++++++++++++++++++++++++++++++++++++++++++++++
\begin{table}
\begin{center}
\caption{Camera Rise/Fall times}
\begin{tabular}{l c}
\hline
Camera	& Rise time (sec) \\
\hline
LWP		& 0.123$\pm$0.004 \\
LWR (at -4.5 kV)& 0.126$\pm$0.006 \\
SWP		& 0.123$\pm$0.005 \\
\hline
\end{tabular}
\label{table:crt}
\end{center}
\end{table}
%+++++++++++++++++++++++++++++++++++++++++++++++++++++++++++++

The actual duration of the shortest exposure times (less than 1 sec. for
the brightest standard stars) is further affected by the Command Decoder
Cycle Time (CDCT) which causes exposure times to be, 2/3 of the times 10.4
msec. longer than t$_{\rm com}$, and the remaining 1/3 is 19.6 msec.
shorter (Oliversen 1987). This effect can be accounted for by taking a
large number of spectra of the same star and comparing each individual
observation with the average spectrum. The mean spectra of the bright
standard stars used for the derivation of the calibration were obtained by
averaging a sufficiently large number of spectra with identical exposure
times and no correction for this effect in the individual exposure times
was necessary.

\subsubsection{Correction for temperature dependence}

The sensitivity of the IUE cameras depends substantially on the temperature
of the Camera Head Amplifier (THDA):

\begin{equation}
\rm{FN_{cor}=\frac{FN_{obs}}{1+C\times(THDA-T_{ref})}}
\end{equation}

\noindent where C represents the change in sensitivity introduced 
by a departure of one degree from the reference temperature, T$_{ref}$,
(e.g. a difference of 5 degrees from T$_{ref}$ represents a 2.5\%\
sensitivity variation in the SWP camera).  We have adopted the parameters
given by Garhart (1991) to correct for this effect (Table \ref{tab:thda}).

%+++++++++++++++++++++++++++++++++++++++++++++++++++++++++++++
\begin{table}
\begin{center}
\caption{THDA dependence parameters}
\begin{tabular}{l c c}
\hline
Camera	&	T$_{\rm ref}$	&	C  \\
\hline
LWP	&	9.5	& -0.0046$\pm$0.0003	\\
LWR	&	14.5	& -0.0088$\pm$0.0004	\\
SWP	&	9.4	& -0.0019$\pm$0.0003	\\
\hline
\end{tabular}
\label{tab:thda}
\end{center}
\end{table}
%+++++++++++++++++++++++++++++++++++++++++++++++++++++++++++++

\subsubsection{The Time Sensitivity Degradation Correction algorithm}

The zero epoch of the IUE calibration was defined to be 1985.0, because at
this time the higher quality ITF observations were performed.  This epoch
was also taken as reference to correct for the loss of sensitivity of the
IUE detectors. The procedure to derive the time sensitivity correction is
fully described in Garhart et al. (1997). In short, fluxes in steps of 5
\AA\ were derived for several hundreds of spectra of the standard stars
covering the whole spacecraft lifetime and normalized to the average of
spectra obtained near the reference epoch 1985.0.  The ratios were binned
into time steps of six months, and then fitted to polynomials over
different time periods. For the LWP camera there are two approaches: after
1984.5, a linear fit is used. Before that epoch, there are few data
available, and a linear interpolation between each pair of points is used.
For the LWR, a fourth order polynomial is used. For SWP, after 1979.5 a
linear fit is used. Prior to this date the same approach as for the early
LWP data is used. These corrections were all derived from pre--1990
data. The corrections were updated after the end of orbital operations to
avoid the need for extrapolation.

As mentioned above, no correction for time--dependent sensitivity
degradation was needed for the single epoch data used for the derivation of
the flux calibration, and therefore no additional uncertainty was
introduced in the calibration by the time dependent sensitivity correction
algorithm.

\subsection{The Zero Point of the Absolute flux scale}
\label{sec:zero}

The direct use of white dwarf atmospheres to define the absolute flux scale
was discarded {\it a priori} by the IUE Project due to the possible errors
implied in the determination of the stellar parameters. Normalization to
optical photometry and/or spectrophotometry was also excluded to avoid the
extrapolation over a wide spectral range, which could amplify substantially
the errors.

We derived the zero point of the Final Archive absolute flux scale {\it
directly} from {\it ultraviolet} observations.  To this purpose we used as
reference the OAO--2 fluxes in the 2100--2300 \AA\ band.  The reason for
the selection of this window is that in this particular wavelength region
the OAO--2 and the TD1 measurements show the best agreement (Beeckmans
1977).

The procedure used to obtain this scale factor was as follows:

\begin{itemize}

\item
The 1991 calibration data were used to obtain average NET spectra
(background--subtracted spectra in units of FN/s) in the LWP range for the
four bright standard stars $\eta$ Aur, $\lambda$ Lep, 10 Lac and $\zeta$
Dra for (all observed with OAO--2) and for G191--B2B.

\item
A preliminary LWP inverse sensitivity curve for 1991 data was obtained by
dividing the above average LWP NET spectra of G191--B2B by the
corresponding model fluxes (normalized to the optical spectrophotometry of
Massey et al. 1998).  A bi--cubic spline fit through the model fluxes has
been used in this process.

\item
The NET 2100-2300 \AA\ spectra of the four standard stars were first
flux--calibrated using the preliminary inverse sensitivity curve and then
compared with the average OAO--2 flux in the same band, as given in Meade
(1978).

\end{itemize}

The result was that the OAO--2 fluxes of the four standards in the band
2100--2300 \AA\ are on average lower by a factor of 1.042 than the ones
obtained from IUE observations in the G191--B2B scale normalised to the
optical spectrophotometry, as shown in Table~\ref{tab:OAO}.

Therefore, the model fluxes provided by Finley, after normalization to
Massey et al. (1988) optical spectrophotometry, still had to be be divided
by a factor of 1.042 to agree with the OAO--2 absolute flux scale. This
scaling factor defines the UV absolute flux scale of the IUE instruments.

%+++++++++++++++++++++++++++++++++++++++++++++++++++++++++++++
\begin{table}
\begin{center}
\caption{The Zero Point Scale Factor}
\begin{tabular}{l c}
\hline
Star		&	G191--B2B scale/OAO--2 scale  \\
\hline
$\eta$ Aur	&	 1.001$\pm$0.024	\\
$\lambda$ Lep	&	 1.025$\pm$0.029	\\
10 Lac		&	 1.062$\pm$0.024	\\
$\zeta$ Dra	&	 1.078$\pm$0.029	\\
		&	 	\\
Average		&	 1.042$\pm$0.035 (rms)	\\
\hline
\end{tabular}
\label{tab:OAO}
\end{center}
\end{table}
%+++++++++++++++++++++++++++++++++++++++++++++++++++++++++++++

\subsection{The Inverse Sensitivity Curves}

The procedure to derive the {\it relative} Inverse Sensitivity Curves
(hereafter ISCs) of the SW and LW cameras in the low resolution mode was
very similar. It can be summarized in two major steps:

\noindent a) Determination of the absolute fluxes of the IUE standard 
stars from the 1991 data:

\begin{itemize}

\item Determination of a mean spectrum for each standard star from the 1991
observations (in units of Flux Number/sec).

All the spectra were individually inspected, rejecting those presenting any
anomaly. All the exposure times were corrected for OBC quantization, THDA
sensitivity and CRFT (see Section \ref{sec:CRFT}). The mean spectrum was
computed by averaging all the available spectra and weighting each point by
its associated error.

\item Derivation of the 1991 ISC from the model and the mean spectrum 
of G191~B2B.

The WD model was divided by the mean spectrum, and the ISC was derived via
a bi-cubic spline, excluding the region around Lyman $\alpha$ and  the
spurious 1515 \AA\ absorption (de la Pe\~na 1992).  The resulting curve was
resampled in bins of 10 and 15 \AA\ for the SW and LW cameras,
respectively. Finally, the scaling factor of 1.042 (see above) was applied
to the curves.

\item Determination of the fluxes of the standard stars from
the 1991 ISC.

The final ISCs were applied to the mean 1991 spectra of the standard stars
in order to derive their absolute fluxes.  The fluxes so obtained define
the absolute flux scale of IUE data (see Appendix A).

\end{itemize}

\noindent b) Derivation of the ISCs for the 1985 Calibration epoch:

\begin{itemize}

\item Determination of a mean net spectrum for each standard star from the 1985
observations (in units of Flux Number/sec).

\item Determination of the 1985 ISC from the 1985 spectra of the standard
stars and their relative fluxes.

For each of the standard stars, an ISC was computed from the average 1985
spectra and the absolute fluxes of the standard stars derived as explained
above. Average ISC were derived from the OAO and TD1 stars. The OAO curve
was scaled to the TD1 one, and both were averaged weighting by the number
of spectra used in each set. In the case of the LWR camera, separate ISC
were derived for both ITFs.

The ISCs derived following this procedure are only applicable to low
resolution Large Aperture Point spectra. Suitable scaling factors for Small
Aperture and Trailed Low Resolution spectra are given by Garhart et
al. (1997).

\end{itemize}

The procedure to derive the high resolution flux calibration from the low
resolution calibration is described by Cassatella et al. (2000).

%%%%%%%%%%%%%%%%%%%%%%%%%%%%%%%%%%%%%%%%%%%%%%%%%%%%%%%%%%%%%%%%%%%%%
\begin{figure}
\resizebox{\hsize}{!}
{\includegraphics{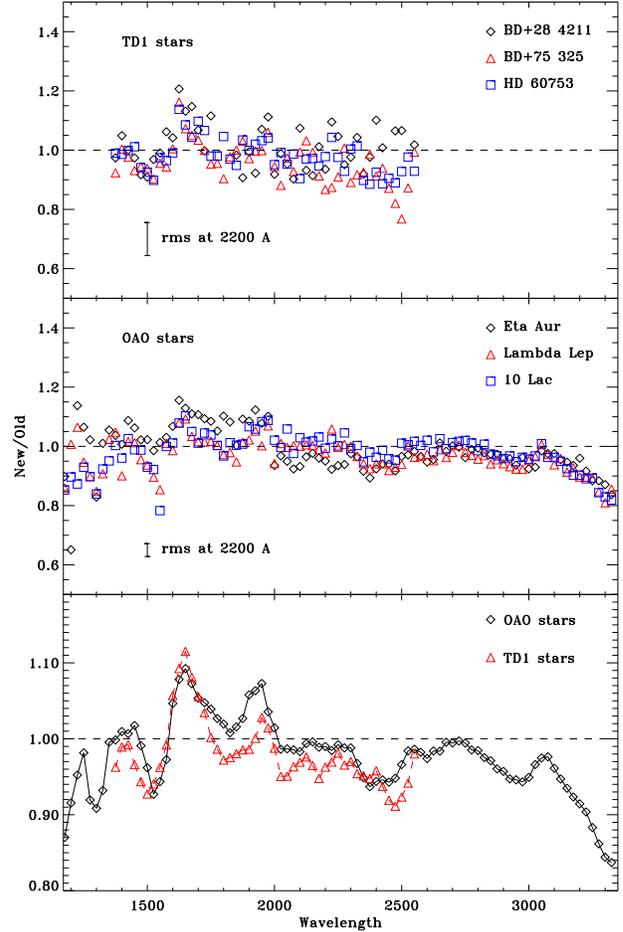}}
\caption{Comparison of the fluxes of the IUE standard stars derived for the
present calibration with those provided by Bohlin and Holm (1984). Shown
are the ratios for stars observed with TD1 (upper panel) and with OAO--2
(middle panel). The bottom panel shows the average ratio for the two groups
of stars.}
\label{fig:ratio}
\end{figure}
%%%%%%%%%%%%%%%%%%%%%%%%%%%%%%%%%%%%%%%%%%%%%%%%%%%%%%%%%%%%%%%%%%%%%

\section{Comparison with other calibrations}

\subsection{Comparison with previous IUE calibrations}

As already pointed out in Sec.~\ref{sec:iueflux}, there was growing
evidence that systematic errors were present in the calibrations prior to
the {\it IUEFA}.  In the following we make a comparison between the present
flux calibration and the previous one by ratioing the fluxes of the
standard stars used to derive them.  The results are shown in
Fig.~\ref{fig:ratio}, where we represent the ratio between the new and old
fluxes for the individual standard stars and, separately, the average
ratios for faint (TD1) and bright (OAO--2) standards.  To compute these
ratios we have used the ``corrected'' fluxes of the standard stars as given
in Bohlin and Holm (1984). We stress that these fluxes are not the original
ones provided by the TD1 and OAO--2 experiments, but are corrected using
the factors derived by these authors to transfer them into the $\eta$ UMa
scale defined by Bohlin et al. (1980). The discontinuities in the flux
ratios shown in the figure clearly indicate the errors in the previous flux
scale. These were most likely introduced by the ``correction factors''
themselves.

%%%%%%%%%%%%%%%%%%%%%%%%%%%%%%%%%%%%%%%%%%%%%%%%%%%%%%%%%%%%%%%%%%%%%
\begin{figure}[t]
\resizebox{\hsize}{!}
{\includegraphics{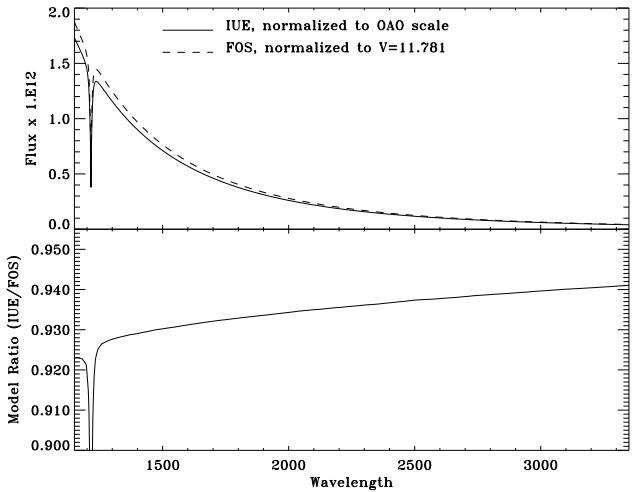}}
\caption{Comparison of the models of the WD G191-B2B used for the
calibration of IUE and HST--FOS. The model used for IUE is normalised as to
agree with the OAO--2 scale at 2200 \AA, i.e.  the flux of the model
originally normalised to the spectrophotometry of Massey et al. (1988) has
been divided by a factor 1.042 (see text).}
\label{fig:model}
\end{figure}
%%%%%%%%%%%%%%%%%%%%%%%%%%%%%%%%%%%%%%%%%%%%%%%%%%%%%%%%%%%%%%%%%%%%%

In the short wavelength range, there are large fluctuations in the ratios
of up to 20\%\ over intervals less than 100 \AA\ wide. The ratio is more
uniform between 2000 and 3100 \AA\, and then decreases abruptly, with the
new fluxes being lower by up to a 15\%. It is interesting to remark that
the largest discrepancies are present in the region 1500--1700 \AA\ where,
probably not accidentally, the differences between TD1 and OAO--2 fluxes
are maximum.

The broad features visible in the flux ratio shown in Fig.~\ref{fig:ratio}
are remarkably similar to those of the ``correction factor'' derived by
Finley et al. (1990) from the comparison of atmosphere models and
observations of DA white dwarfs, in particular shortward 2000 \AA.

\subsection{Comparison with the HST Absolute Flux Scale}
\label{sec:hst}

White dwarf models have been used for the flux calibration in the UV range
of other space experiments.  This is the case of the Hubble Space Telescope
and the Hopkins Ultraviolet Telescope (HUT, Kruk et al. 1997, 1999). In the
following we will compare the IUE and HST-FOS flux calibrations. The
comparison of HUT (both ASTRO--1 and ASTRO--2) and FOS is described in Kruk
et al. (1997, 1999).

%%%%%%%%%%%%%%%%%%%%%%%%%%%%%%%%%%%%%%%%%%%%%%%%%%%%%%%%%%%%%%%%%%%%%
\begin{figure}[b]
\resizebox{\hsize}{!}
{\includegraphics{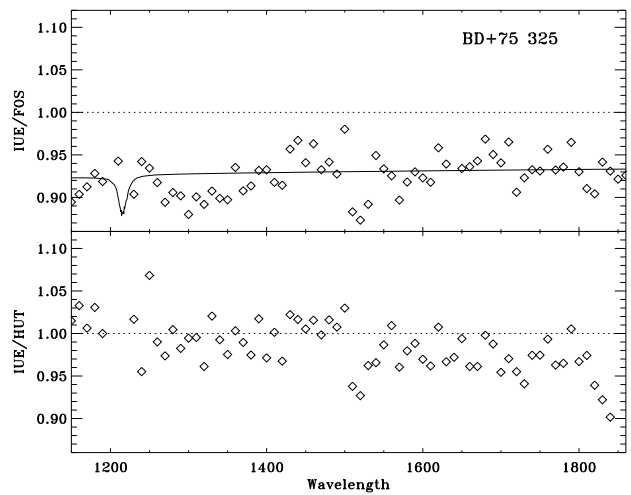}}
\caption{Comparison of the IUE, HST--FOS and HUT fluxes of the standard star
BD+75~325. Shown for comparison in the upper panel is the ratio of the
models of the WD G191--2B2 used for the calibration of IUE and FOS (the
same as shown at the bottom panel of Fig.~\ref{fig:model}). This shows that
although the relative calibrations are quite consistent, the absolute scale
is still rather uncertain.}
\label{fig:hst}
\end{figure}
%%%%%%%%%%%%%%%%%%%%%%%%%%%%%%%%%%%%%%%%%%%%%%%%%%%%%%%%%%%%%%%%%%%%%

The absolute calibration of the HST Faint Object Spectrograph is based on a
slightly different model of the DA WD G191--B2B (Bohlin et al.  1995) with
a pure Hydrogen atmosphere, an effective temperature of 61300~K, log~g=7.5,
and normalised to V=11.781 (Colina and Bohlin 1994). 

The difference in effective temperature of the models used for the
calibrations of IUE and FOS results in a slightly different slope in the UV
range (approximately 1\%, see Fig.~\ref{fig:model}).  The model fluxes used
for the IUE calibration (with the original scaling to optical
spectrophotometry) are lower than the model used for FOS by 1.1 \%\ at 5500
\AA\ due to the different normalisation.  The additional 4.2\%\ scaling
factor makes this difference 5.3\%\ at V (in the sense that the FOS model
is brighter. This normalisation implies a V magnitude of 11.84 for
G191--B2B, in contrast with the recent revision by Bohlin (2000) which
derived a value of V=11.773$\pm$0.0012(1 $\sigma$).

The slightly different slope of the models increases this
discrepancy in the IUE range. The average ratio of the models used in the
IUE and the FOS calibrations in the range 1150-3350 \AA\ (excluding the
region around Lyman $\alpha$) is 0.933, i.e.  model used for the IUE
calibration is lower by a 7.2\%.

We have compared the IUE, FOS (Bohlin 1996) and HUT (Kruk, private
communication) absolute fluxes of the standard star and BD+75~325 in the
spectral region of overlap of the three experiments, as shown in
Fig.~\ref{fig:hst}.  The continuous line in the upper panel of the figure
represents the ratio between the models used for the IUE and FOS
calibrations. The average ratios IUE/other over the common wavelength range
is 0.93$\pm$0.03 and 0.99$\pm$0.05 for FOS and HUT, respectively.  The
figure shows that the overall agreement between IUE and FOS flux and model
ratios is good, although there are some broad features, which are thought
to be induced by the effects of the residual non--linearities of the IUE
cameras on the spectra used for the calibration.  On average, the flux
ratio IUE/FOS is within a 3\%\ of the model ratio, except for the region
2250-2450 \AA\, where it is lower by a 4\%.

The IUE fluxes seem to agree better with the HUT data, but this might
simply be an accidental artifact, due to the complex calibration of the
different HUT instrumental configurations and the large uncertainties
involved (Kruk, private communication).

Although the comparisons in Fig. \ref{fig:hst} show a general agreement in
the three calibrations (IUE independent of HST and HUT, which are based on
the same absolute fluxes), to $\pm$3\% in the relative calibration, it is
clear that the absolute UV scale is still uncertain to $\pm$10\%.

%%%%%%%%%%%%%%%%%%%%%%%%%%%%%%%%%%%%%%%%%%%%%%%%%%%%%%%%%%%%%%%%%%%%%
\begin{figure}
\resizebox{\hsize}{!}
{\includegraphics{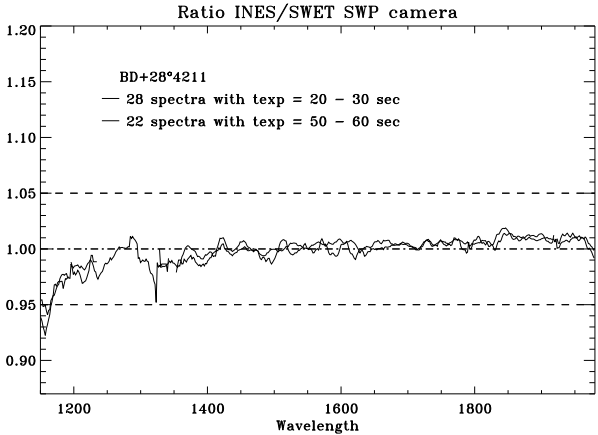}}
\caption{Average ratio of {\it NEWSIPS} and {\it INES} fluxes for SWP
spectra of the standard star BD+28~4211 (28 spectra with exposure times
between 20 and 30 sec. and 22 spectra with exposure times between 50 and 60
sec.). The thin line corresponds to the average ratio for non--saturated
spectra, and the thick one to saturated spectra, with only the
non--saturated region shown.  The dashed lines mark the $\pm$ 5\%\ limits.}
\label{fig:iness}
\end{figure}
%%%%%%%%%%%%%%%%%%%%%%%%%%%%%%%%%%%%%%%%%%%%%%%%%%%%%%%%%%%%%%%%%%%%%

\subsection{Applicability to {\it INES}--extracted data}
\label{sec:ines}

The flux calibration described in this paper has been derived from IUE low
resolution spectra processed with {\it NEWSIPS} and the {\it SWET} optimal
extraction procedure (Garhart et al. 1997). IUE low resolution data have
been re-extracted from the {\it NEWSIPS} bi-dimensional SILO files using a
different algorithm for the {\it INES} archive.  The {\it INES} extraction
is described in detail by Rodr\'{\i}guez-Pascual et al. (1999).  It
includes, among other features, new noise models and an improved extraction
procedure.  Both extraction algorithms ({\it SWET} and {\it INES}) use the
same inverse sensitivity curve, therefore any differences in the flux
calibrated spectra would also appear in NET spectra. Differences of this kind
could in principle arise from the different procedures used to estimate the
background level, to evaluate the spatial profile, and from the adopted
noise model.

%%%%%%%%%%%%%%%%%%%%%%%%%%%%%%%%%%%%%%%%%%%%%%%%%%%%%%%%%%%%%%%%%%%%%
\begin{figure}[b]
\resizebox{\hsize}{!}
{\includegraphics{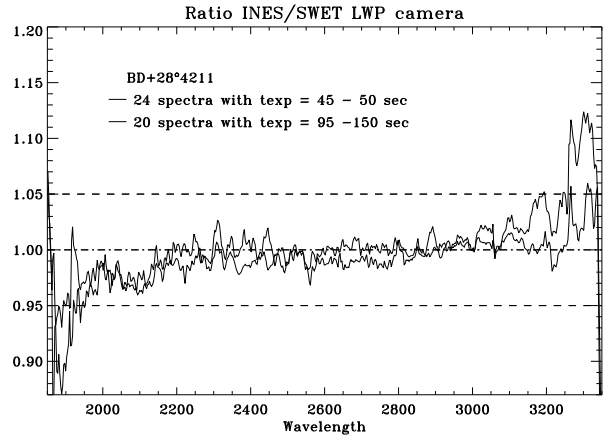}}
\caption{Same as Fig.~\ref{fig:iness}, but for the LWP camera.  Shown are
the average ratios corresponding to 24 spectra with exposure times between
45 and 50 sec. and 20 spectra with exposure times between 95 and 150
sec. of the standard star BD+28~4211.}
\label{fig:inesl}
\end{figure}
%%%%%%%%%%%%%%%%%%%%%%%%%%%%%%%%%%%%%%%%%%%%%%%%%%%%%%%%%%%%%%%%%%%%%

\begin{figure}
\resizebox{\hsize}{!}
{\includegraphics{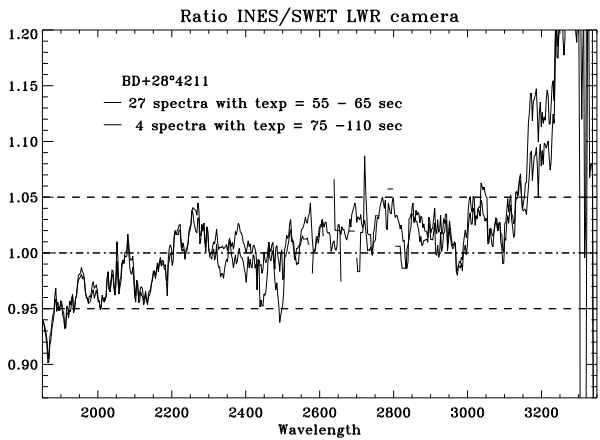}}
\caption{Same as Fig.~\ref{fig:iness}, but for the LWR camera.  Shown are
the average ratios for 27 spectra with exposure times between 55 and 65
sec. and 4 spectra with exposure times between 75 and 110 sec. of
BD+28~4211.  All the spectra have been processed with ITF--B (see Section
\ref{sec:itf}) and were taken in the period during which the camera was
still operational (1978--1983).}
\label{fig:inesr}
\end{figure}

In order to check the applicability of the Final Archive calibration to
{\it INES}--extracted data we have taken low resolution spectra of the IUE
standard star BD+28~4211 and compared the {\it INES} and the {\it NEWSIPS}
fluxes. For this purpose we have divided the spectra into two groups
according to their level of exposure: the first group containing
non--saturated spectra, and the second one containing spectra saturated in
the region of maximum sensitivity of the cameras. We have computed the mean
ratio for each group of spectra. The results are shown in
Fig.~\ref{fig:iness}, Fig.~\ref{fig:inesl} and Fig.~\ref{fig:inesr}.

In the SWP camera, the average flux ratio {\it INES}/{\it SWET} for the
short--exposure time spectra is 1.00$\pm$0.01 longward 1250 \AA, with a
slight slope along the full wavelength range (i.e. the {\it INES} flux is
slightly lower than the {\it SWET} flux shortward 1400 \AA\ and slightly
higher longward 1600 \AA). Shortward of Lyman $\alpha$ the {\it INES} flux
is up to a 8 \% lower. The {\it INES} flux is also lower in this spectral
range for the longest exposure spectra, but only by less than 4\%.
Longward 1400 \AA\ the flux ratio is 1.00$\pm$0.01 independently on the
level of exposure of the spectra.

In the case of the LWP camera, the {\it INES} and {\it SWET} fluxes agree
within 1\%\ along most of the spectral range (2200-3000 \AA). The largest
differences are found at the edges of the range. While at the short
wavelengths the {\it INES} extraction provides fluxes up to a 10\%\ lower
than {\it SWET}, the contrary occurs longward 3000 \AA\, where {\it INES}
fluxes can be a 10\%\ higher. It must be noted that in both cases the
differences are larger for short--exposure time spectra, suggesting that
the discrepancy can originate from non--linearity effects at low exposure
levels.

The largest discrepancies are found in the LWR camera (for images processed
with ITF--B). In the region 2500--3000 \AA\ the ratio {\it INES}/{\it SWET}
is 1.02$\pm$0.04, while at shorter wavelengths (2100--2500 \AA) it is
closer to unity: 1.01$\pm$0.04.  As in the case of the LWP camera, it is
longward 3200 \AA\ where the difference becomes larger, with the {\it INES}
flux larger by up to a 20\%. In this camera there is no significative
difference in the behaviour of short and long exposure time spectra.

In summary, the differences between the {\it INES} and {\it SWET}
extractions are within a 2\%\ over most of the spectra range, with largest
differences at the edges of the cameras (in the SWP only at the shortest
wavelengths).

%%%%%%%%%%%%%%%%%%%%%%%%%%%%%%%%%%%%%%%%%%%%%%%%%%%%%%%%%%%%%%%%%%%%%
%%%%%%%%%%%%%%%%%%%%%%%%%%%%%%%%%%%%%%%%%%%%%%%%%%%%%%%%%%%%%%%%%%%%%

\section{Conclusions}
\label{sec:conclusions}

In this paper we have described the definition of the flux scale which has
been adopted for the flux calibration of the IUE Final Archive data.  After
having discussed the inadequacy of the $\eta$ UMa flux scale as in Bohlin
et al. (1980), we have shown that a more pertinent method which optimizes
the internal and external consistency of IUE fluxes is to use the DA white
dwarf G191--B2B as primary standard star.  The procedure followed consisted
basically in using the IUE observations of this star obtained in 1991
together with model atmosphere fluxes normalized to the data from optical
spectrophotometry in Massey et al. (1988) to obtain the {\it shape} of the
inverse sensitivity curves for the three IUE operational cameras.  At this
point, the many IUE observations of $\eta$ Aur, $\lambda$ Lep, 10 Lac and
$\zeta$ Dra, also obtained in 1991, were used to find a suitable {\it
scaling factor} of the sensitivity curves such that, after calibration, the
scaled fluxes best fitted the corresponding OAO--2 original measurements
from Meade (1978) in the range 2100--2300 \AA.  The choice of this
wavelength interval to set the scaling factor was to link the IUE
calibration to {\it ultraviolet} instead of optical observations. Also, the
2100--2300 \AA\ range is the one where the agreement between TD1 and OAO--2
fluxes is best.  As shown in Sect. \ref{sec:zero}, the 2100-2300
\AA\ OAO--2 fluxes of the quoted four standards are on average a factor of
1.042 lower than those from the G191--B2B model normalised to the data from
optical spectrophotometry (Massey et al. 1988).

The sensitivity curves for the 1991 epoch, together with the very many
observations of the standard stars secured in this epoch were to define the
absolute fluxes of the standard stars.  These fluxes were then used as
input to derive the sensitivity curves for the 1985 calibration epoch
(i.e. the epoch the ITFs were obtained for the IUE cameras).

The absolute fluxes of six IUE standard stars and the model fluxes of
G191--B2B {\it in the OAO--2 scale} are given in Appendices A and B,
respectively.

As shown in Sec. \ref{sec:hst}, the fluxes obtained with this calibration
are on average 7.2\%\ lower than the ones provided by the Faint Object
Spectrograph on board the Hubble Space Telescope in the range 1150--3350
\AA. This discrepancy can be ascribed to the different choice for scaling
the G191--B2B model fluxes and, to a minor extent, to the slightly different
stellar parameters adopted for the G191--B2B model.

Rodr\'{\i}guez-Pascual et al. (1999) have discussed the {\it INES} system
and its advantages over {\it NEWSIPS} to remove the systematic errors found
in this latter package. In view of the different extraction software used
in the two systems, a specific test has been carried out in this paper to
verify the applicability of the present flux scale to low resolution data
processed with {\it INES}.  The conclusion is that, in spite of the
different extraction algorithms used, the application of the present
calibration to {\it INES} spectra is fully justified.  We stress that the
present paper has a direct application to the absolute calibration of low
resolution spectra. The method used to obtain absolute calibration of high
resolution spectra has been discussed elsewhere (Cassatella et al. 2000).

\begin{acknowledgements}

We would like to thank the attendees of a Meeting on the IUE flux
calibration which we organized at the ESA IUE ground station of Villafranca
del Castillo in December 1990, and in particular, Prof. V. Weidemann and
Drs. D. Finley, D. Husfeld, D. Koester and D. Massa. We are grateful to
Dr. J. Kruk for his comments about the HUT calibration. We would also like
to acknowledge the large effort made by the NASA and ESA IUE Observatories
in supporting the calibration activities. This paper is dedicated to the
memory of Andy Michalitsianos, who has strongly contributed to the success
of the IUE project.

\end{acknowledgements}

\appendix

\section{The Absolute Fluxes of the IUE Standard Stars}

The tables and figures in Appendix A show the absolute fluxes of the IUE
Standard Stars used for the derivation of the cameras Inverse Sensitivity
Curves. These fluxes have been derived as described in the text, i.e. the
{\it relative fluxes} with the model of the WD G191~B2B, and the zero point
of the scale set by OAO-2 observations.  These fluxes define therefore the
absolute flux scale of IUE.  In all cases the wavelength is in \AA\ and the
flux in erg~cm$^{\rm -2}$~s$^{\rm -1}$~\AA$^{\rm -1}$. Note that in some
cases there are gaps in the data, due to the presence of instrumental
artifacts that preclude the accurate determination of the flux in that
wavelength bin.

\begin{table*}
\caption{BD+28~4211} 
\vspace{0.3cm}

\begin{tabular}{c c c c c c c c c c}
\hline
 Wavelength &  Flux &  Wavelength &  Flux &  Wavelength &  Flux &  Wavelength &  Flux &  Wavelength &  Flux   \\
\hline                      
   1150  &  6.27E-11 &   1500  &  2.64E-11 &    1850  &  1.21E-11 &    2315  & 5.03E-12  &     2840  & 2.41E-12 \\
   1160  &  6.25E-11 &   1510  &  2.47E-11 &    1860  &  1.19E-11 &    2330  & 5.13E-12  &     2855  & 2.45E-12 \\  
   1170  &  5.95E-11 &   1520  &  2.41E-11 &    1870  &  1.17E-11 &    2345  & 4.55E-12  &     2870  & 2.34E-12 \\  
   1180  &  5.90E-11 &   1530  &  2.44E-11 &    1880  &  1.15E-11 &    2360  & 4.53E-12  &     2885  & 2.37E-12 \\  
   1190  &  5.43E-11 &   1540  &  2.41E-11 &    1890  &  1.13E-11 &    2375  & 4.61E-12  &     2900  & 2.27E-12 \\  
   1200  &  5.48E-11 &   1550  &  2.26E-11 &    1900  &  1.10E-11 &    2390  & 4.18E-12  &     2915  & 2.20E-12 \\ 
   1210  &  3.59E-11 &   1560  &  2.28E-11 &    1910  &  1.09E-11 &    2405  & 4.37E-12  &     2930  & 2.20E-12 \\  
   1220  &  3.59E-11 &   1570  &  2.26E-11 &    1920  &           &    2420  & 4.38E-12  &     2945  & 2.14E-12 \\  
   1230  &  4.86E-11 &   1580  &  2.18E-11 &    1930  &           &    2435  & 4.25E-12  &     2960  & 2.10E-12 \\  
   1240  &  4.33E-11 &   1590  &  2.10E-11 &    1940  &  1.01E-11 &    2450  & 4.35E-12  &     2975  & 2.10E-12 \\  
   1250  &  4.80E-11 &   1600  &  2.04E-11 &    1950  &  9.94E-12 &    2465  & 4.06E-12  &     2990  & 2.02E-12 \\  
   1260  &  4.70E-11 &   1610  &  1.00E-11 &    1960  &  9.82E-12 &    2480  & 3.93E-12  &     3005  & 2.00E-12 \\  
   1270  &  4.41E-11 &   1620  &  1.99E-11 &    1970  &  9.67E-12 &    2495  & 3.98E-12  &     3020  & 1.95E-12 \\  
   1280  &  4.41E-11 &   1630  &  1.87E-11 &    1980  &  9.62E-12 &    2510  & 3.68E-12  &     3035  & 1.97E-12 \\  
   1290  &  4.43E-11 &   1640  &  1.68E-11 &    2000  & 8.60E-12  &    2525  & 3.78E-12  &     3050  & 1.68E-12 \\  
   1300  &  4.21E-11 &   1650  &  1.83E-11 &    2015  & 8.48E-12  &    2540  & 3.80E-12  &     3065  & 1.88E-12 \\  
   1310  &  4.09E-11 &   1660  &  1.82E-11 &    2030  & 8.23E-12  &    2555  & 3.68E-12  &     3080  & 1.83E-12 \\  
   1320  &  3.90E-11 &   1670  &  1.79E-11 &    2045  & 8.23E-12  &    2570  & 3.62E-12  &     3095  & 1.83E-12 \\  
   1330  &  4.05E-11 &   1680  &  1.77E-11 &    2060  & 8.10E-12  &    2585  & 3.55E-12  &     3110  & 1.80E-12 \\  
   1340  &  3.68E-11 &   1690  &  1.71E-11 &    2075  & 7.44E-12  &    2600  & 3.55E-12  &     3125  & 1.77E-12 \\  
   1350  &  3.68E-11 &   1700  &  1.64E-11 &    2090  & 7.82E-12  &    2615  & 3.50E-12  &     3140  & 1.68E-12 \\  
   1360  &  3.65E-11 &   1710  &  1.58E-11 &    2105  & 7.50E-12  &    2630  & 3.30E-12  &     3155  & 1.72E-12 \\  
   1370  &  3.42E-11 &   1720  &  1.57E-11 &    2120  & 7.19E-12  &    2645  & 3.28E-12  &     3170  & 1.66E-12 \\  
   1380  &  3.57E-11 &   1730  &  1.50E-11 &    2135  & 7.36E-12  &    2660  & 3.25E-12  &     3185  & 1.60E-12 \\  
   1390  &  3.43E-11 &   1740  &  1.52E-11 &    2150  & 6.51E-12  &    2675  & 3.14E-12  &     3200  & 1.47E-12 \\  
   1400  &  3.33E-11 &   1750  &  1.50E-11 &    2165  & 6.51E-12  &    2690  & 3.04E-12  &     3215  & 1.52E-12 \\  
   1410  &  3.23E-11 &   1760  &  1.49E-11 &    2180  & 6.20E-12  &    2705  & 3.08E-12  &     3230  & 1.49E-12 \\  
   1420  &  3.13E-11 &   1770  &  1.43E-11 &    2195  & 6.19E-12  &    2720  & 2.89E-12  &     3245  & 1.49E-12 \\  
   1430  &  3.20E-11 &   1780  &  1.42E-11 &    2210  & 5.55E-12  &    2735  & 2.64E-12  &     3260  & 1.46E-12 \\  
   1440  &  3.03E-11 &   1790  &  1.41E-11 &    2225  & 5.96E-12  &    2750  & 2.89E-12  &     3275  & 1.44E-12 \\  
   1450  &  2.96E-11 &   1800  &  1.34E-11 &    2240  & 5.50E-12  &    2765  & 2.70E-12  &     3290  & 1.35E-12 \\  
   1460  &  2.89E-11 &   1810  &  1.30E-11 &    2255  & 5.60E-12  &    2780  & 2.74E-12  &     3305  & 1.39E-12 \\  
   1470  &  2.78E-11 &   1820  &  1.28E-11 &    2270  & 5.69E-12  &    2795  & 2.61E-12  &     3320  & 1.35E-12 \\  
   1480  &  2.74E-11 &   1830  &  1.27E-11 &    2285  & 5.26E-12  &    2810  & 2.58E-12  &     3335  & 1.37E-12 \\  
   1490  &  2.64E-11 &   1840  &  1.23E-11 &    2300  & 5.33E-12  &    2825  & 2.60E-12  &     3350  & 1.41E-12 \\  
\hline
\end{tabular}
\end{table*}

\begin{table*}
\caption{BD+75~325} 
\vspace{0.3cm}
\begin{tabular}{c c c c c c c c c c}
\hline
 Wavelength &  Flux &  Wavelength &  Flux &  Wavelength &  Flux &  Wavelength &  Flux &  Wavelength &  Flux   \\
\hline
   1150  &  9.01E-11 & 1500  &  4.26E-11 & 1850  &    2.36E-11 & 2315  &    1.13E-11 & 2840  &    5.95E-12  \\         1160  &  9.20E-11 & 1510  &  4.11E-11 & 1860  &    2.36E-11 & 2330  &    1.22E-11 & 2855  &    5.86E-12  \\      
   1170  &  8.43E-11 & 1520  &  3.93E-11 & 1870  &    2.34E-11 & 2345  &    1.13E-11 & 2870  &    5.60E-12  \\     
   1180  &  8.11E-11 & 1530  &  3.86E-11 & 1880  &    2.29E-11 & 2360  &    1.12E-11 & 2885  &    5.57E-12  \\      
   1190  &  7.56E-11 & 1540  &  4.03E-11 & 1890  &    2.28E-11 & 2375  &    1.07E-11 & 2900  &    5.39E-12  \\      
   1200  &  7.94E-11 & 1550  &  3.80E-11 & 1900  &    2.22E-11 & 2390  &    9.49E-12 & 2915  &    5.38E-12  \\         1210  &  6.06E-11 & 1560  &  3.76E-11 & 1910  &    2.22E-11 & 2405  &    1.00E-11 & 2930  &    5.39E-12  \\      
   1220  &  5.88E-11 & 1570  &  3.56E-11 & 1920  &             & 2420  &    1.03E-11 & 2945  &    5.15E-12  \\      
   1230  &  6.76E-11 & 1580  &  3.67E-11 & 1930  &             & 2435  &    1.01E-11 & 2960  &    5.00E-12  \\      
   1240  &  5.54E-11 & 1590  &  3.62E-11 & 1940  &    2.09E-11 & 2450  &    1.00E-11 & 2975  &    5.05E-12  \\      
   1250  &  6.27E-11 & 1600  &  3.48E-11 & 1950  &    2.06E-11 & 2465  &    9.82E-12 & 2990  &    4.91E-12  \\      
   1260  &  6.20E-11 & 1610  &  3.27E-11 & 1960  &    2.00E-11 & 2480  &    9.49E-12 & 3005  &    4.81E-12  \\      
   1270  &  6.17E-11 & 1620  &  3.34E-11 & 1970  &    1.99E-11 & 2495  &    9.40E-12 & 3020  &    4.69E-12  \\      
   1280  &  6.22E-11 & 1630  &  3.17E-11 & 1980  &    1.99E-11 & 2510  &    8.02E-12 & 3035  &    4.76E-12  \\      
   1290  &  6.50E-11 & 1640  &  2.98E-11 & 2000  &    1.85E-11 & 2525  &    8.72E-12 & 3050  &    4.47E-12  \\      
   1300  &  6.06E-11 & 1650  &  3.22E-11 & 2015  &    1.87E-11 & 2540  &    9.08E-12 & 3065  &    4.53E-12  \\      
   1310  &  5.71E-11 & 1660  &  3.22E-11 & 2030  &    1.81E-11 & 2555  &    8.64E-12 & 3080  &    4.38E-12  \\      
   1320  &  5.58E-11 & 1670  &  3.23E-11 & 2045  &    1.83E-11 & 2570  &    8.49E-12 & 3095  &    4.38E-12  \\      
   1330  &  5.84E-11 & 1680  &  3.31E-11 & 2060  &    1.71E-11 & 2585  &    8.26E-12 & 3110  &    4.36E-12  \\      
   1340  &  5.53E-11 & 1690  &  3.21E-11 & 2075  &    1.64E-11 & 2600  &    8.11E-12 & 3125  &    4.24E-12  \\      
   1350  &  5.41E-11 & 1700  &  3.03E-11 & 2090  &    1.64E-11 & 2615  &    8.07E-12 & 3140  &    4.11E-12  \\      
   1360  &  5.24E-11 & 1710  &  2.96E-11 & 2105  &    1.62E-11 & 2630  &    7.73E-12 & 3155  &    4.10E-12  \\      
   1370  &  5.20E-11 & 1720  &  2.72E-11 & 2120  &    1.53E-11 & 2645  &    7.55E-12 & 3170  &    4.03E-12  \\      
   1380  &  5.16E-11 & 1730  &  2.78E-11 & 2135  &    1.50E-11 & 2660  &    7.47E-12 & 3185  &    3.78E-12  \\      
   1390  &  5.12E-11 & 1740  &  2.89E-11 & 2150  &    1.50E-11 & 2675  &    7.45E-12 & 3200  &    3.26E-12  \\      
   1400  &  5.02E-11 & 1750  &  2.81E-11 & 2165  &    1.41E-11 & 2690  &    7.21E-12 & 3215  &    3.57E-12  \\      
   1410  &  4.88E-11 & 1760  &  2.85E-11 & 2180  &    1.43E-11 & 2705  &    7.25E-12 & 3230  &    3.59E-12  \\      
   1420  &  4.77E-11 & 1770  &  2.76E-11 & 2195  &    1.32E-11 & 2720  &    6.80E-12 & 3245  &    3.63E-12  \\      
   1430  &  4.84E-11 & 1780  &  2.75E-11 & 2210  &    1.34E-11 & 2735  &    5.76E-12 & 3260  &    3.51E-12  \\      
   1440  &  4.77E-11 & 1790  &  2.76E-11 & 2225  &    1.33E-11 & 2750  &    6.72E-12 & 3275  &    3.29E-12  \\      
   1450  &  4.41E-11 & 1800  &  2.58E-11 & 2240  &    1.37E-11 & 2765  &    6.54E-12 & 3290  &    3.27E-12  \\      
   1460  &  4.46E-11 & 1810  &  2.48E-11 & 2255  &    1.17E-11 & 2780  &    6.51E-12 & 3305  &    3.17E-12  \\      
   1470  &  4.40E-11 & 1820  &  2.45E-11 & 2270  &    1.28E-11 & 2795  &    6.19E-12 & 3320  &    3.14E-12  \\      
   1480  &  4.33E-11 & 1830  &  2.52E-11 & 2285  &    1.25E-11 & 2810  &    6.13E-12 & 3335  &    3.27E-12  \\      
   1490  &  4.20E-11 & 1840  &  2.41E-11 & 2300  &    1.20E-11 & 2825  &    6.14E-12 & 3350  &    3.61E-12  \\
\hline
\end{tabular}
\end{table*}

\begin{table*}
\caption{HD~60753} 
\vspace{0.3cm}

\begin{tabular}{c c c c c c c c c c}
\hline
 Wavelength &  Flux &  Wavelength &  Flux &  Wavelength &  Flux &  Wavelength &  Flux &  Wavelength &  Flux   \\
\hline
   1150  &   8.01E-11  & 1500  &   7.49E-11   & 1850  &   4.69E-11  & 2315  &   3.15E-11  & 2840  &   2.25E-11  \\  
   1160  &   8.22E-11  & 1510  &   6.91E-11   & 1860  &   4.69E-11  & 2330  &   3.17E-11  & 2855  &   2.24E-11  \\ 
   1170  &   8.21E-11  & 1520  &   6.86E-11   & 1870  &   5.09E-11  & 2345  &   2.74E-11  & 2870  &   2.29E-11  \\ 
   1180  &   7.54E-11  & 1530  &   6.58E-11   & 1880  &   4.87E-11  & 2360  &   2.83E-11  & 2885  &   2.22E-11  \\ 
   1190  &   7.61E-11  & 1540  &   6.68E-11   & 1890  &   4.50E-11  & 2375  &   2.76E-11  & 2900  &   2.20E-11  \\ 
   1200  &   3.47E-11  & 1550  &   6.65E-11   & 1900  &   4.60E-11  & 2390  &   2.77E-11  & 2915  &   2.19E-11  \\ 
   1210  &   7.63E-12  & 1560  &   6.49E-11   & 1910  &   4.69E-11  & 2405  &   2.75E-11  & 2930  &   2.10E-11  \\ 
   1220  &   1.74E-11  & 1570  &   6.53E-11   & 1920  &             & 2420  &   2.87E-11  & 2945  &   2.10E-11  \\ 
   1230  &   6.33E-11  & 1580  &   6.67E-11   & 1930  &             & 2435  &   2.71E-11  & 2960  &   2.06E-11  \\ 
   1240  &   9.12E-11  & 1590  &   6.78E-11   & 1940  &   4.49E-11  & 2450  &   2.86E-11  & 2975  &   2.06E-11  \\ 
   1250  &   9.32E-11  & 1600  &   6.45E-11   & 1950  &   4.39E-11  & 2465  &   2.70E-11  & 2990  &   2.09E-11  \\ 
   1260  &   8.54E-11  & 1610  &   6.26E-11   & 1960  &   4.23E-11  & 2480  &   2.66E-11  & 3005  &   2.02E-11  \\ 
   1270  &   9.19E-11  & 1620  &   6.87E-11   & 1970  &   4.49E-11  & 2495  &   2.80E-11  & 3020  &   2.00E-11  \\ 
   1280  &   9.44E-11  & 1630  &   6.69E-11   & 1980  &   4.51E-11  & 2510  &   2.81E-11  & 3035  &   2.08E-11  \\ 
   1290  &   8.93E-11  & 1640  &   6.69E-11   & 2000  &   3.95E-11  & 2525  &   2.78E-11  & 3050  &   1.74E-11  \\ 
   1300  &   6.62E-11  & 1650  &   6.66E-11   & 2015  &   4.26E-11  & 2540  &   2.64E-11  & 3065  &   1.97E-11  \\ 
   1310  &   8.16E-11  & 1660  &   6.37E-11   & 2030  &   3.98E-11  & 2555  &   2.65E-11  & 3080  &   1.98E-11  \\ 
   1320  &   9.69E-11  & 1670  &   6.39E-11   & 2045  &   3.92E-11  & 2570  &   2.69E-11  & 3095  &   1.99E-11  \\ 
   1330  &   8.11E-11  & 1680  &   6.57E-11   & 2060  &   3.87E-11  & 2585  &   2.62E-11  & 3110  &   1.96E-11  \\ 
   1340  &   8.44E-11  & 1690  &   6.48E-11   & 2075  &   3.86E-11  & 2600  &   2.63E-11  & 3125  &   1.95E-11  \\ 
   1350  &   9.39E-11  & 1700  &   6.33E-11   & 2090  &   3.68E-11  & 2615  &   2.64E-11  & 3140  &   1.86E-11  \\ 
   1360  &   9.25E-11  & 1710  &   5.83E-11   & 2105  &   3.61E-11  & 2630  &   2.63E-11  & 3155  &   1.89E-11  \\ 
   1370  &   8.76E-11  & 1720  &   5.55E-11   & 2120  &   3.72E-11  & 2645  &   2.71E-11  & 3170  &   1.83E-11  \\ 
   1380  &   8.91E-11  & 1730  &   5.67E-11   & 2135  &   3.50E-11  & 2660  &   2.62E-11  & 3185  &   1.75E-11  \\ 
   1390  &   8.30E-11  & 1740  &   5.75E-11   & 2150  &   3.45E-11  & 2675  &   2.60E-11  & 3200  &   1.72E-11  \\ 
   1400  &   8.05E-11  & 1750  &   5.70E-11   & 2165  &   3.45E-11  & 2690  &   2.56E-11  & 3215  &   1.65E-11  \\ 
   1410  &   8.26E-11  & 1760  &   5.77E-11   & 2180  &   3.12E-11  & 2705  &   2.57E-11  & 3230  &   1.70E-11  \\ 
   1420  &   8.20E-11  & 1770  &   5.60E-11   & 2195  &   3.21E-11  & 2720  &   2.51E-11  & 3245  &   1.72E-11  \\ 
   1430  &   8.36E-11  & 1780  &   5.62E-11   & 2210  &   3.20E-11  & 2735  &   2.52E-11  & 3260  &   1.72E-11  \\ 
   1440  &   8.35E-11  & 1790  &              & 2225  &   3.30E-11  & 2750  &   2.42E-11  & 3275  &   1.67E-11  \\ 
   1450  &   8.69E-11  & 1800  &   5.73E-11   & 2240  &   3.13E-11  & 2765  &   2.38E-11  & 3290  &   1.65E-11  \\ 
   1460  &   8.57E-11  & 1810  &   5.39E-11   & 2255  &   3.09E-11  & 2780  &   2.47E-11  & 3305  &   1.59E-11  \\ 
   1470  &   8.03E-11  & 1820  &   5.28E-11   & 2270  &   3.10E-11  & 2795  &   2.31E-11  & 3320  &   1.43E-11  \\ 
   1480  &   7.74E-11  & 1830  &   5.34E-11   & 2285  &   3.08E-11  & 2810  &   2.33E-11  & 3335  &   1.55E-11  \\ 
   1490  &   7.70E-11  & 1840  &   5.10E-11   & 2300  &   3.19E-11  & 2825  &   2.39E-11  & 3350  &   1.51E-11  \\ 
\hline
\end{tabular}
\end{table*}

\begin{table*}
\caption{HD~32630} 
\vspace{0.3cm}

\begin{tabular}{c c c c c c c c c c}
\hline
 Wavelength &  Flux &  Wavelength &  Flux &  Wavelength &  Flux &  Wavelength &  Flux &  Wavelength &  Flux   \\
\hline
   1150  & 4.22E-09  & 1500  & 3.11E-09 &   1850  &    1.86E-09 &  2315  &    1.14E-09  &  2840  &    7.20E-10  \\
   1160  & 4.23E-09 & 1510  &  2.74E-09 &   1860  &    1.90E-09 &  2330  &    1.09E-09  &  2855  &    7.27E-10	 \\
   1170  & 3.95E-09 & 1520  &  2.91E-09 &   1870  &    1.99E-09 &  2345  &    9.91E-10  &  2870  &    7.09E-10	 \\
   1180  & 3.97E-09 & 1530  &  2.66E-09 &   1880  &    1.94E-09 &  2360  &    1.01E-09  &  2885  &    6.95E-10	 \\
   1190  & 3.79E-09 & 1540  &  2.77E-09 &   1890  &    1.81E-09 &  2375  &    9.56E-10  &  2900  &    6.74E-10	 \\
   1200  & 1.94E-09 & 1550  &  2.76E-09 &   1900  &    1.85E-09 &  2390  &    9.67E-10  &  2915  &    6.71E-10	 \\
   1210  & 6.41E-10 & 1560  &  2.65E-09 &   1910  &    1.85E-09 &  2405  &    9.72E-10  &  2930  &    6.59E-10	 \\
   1220  & 1.00E-09 & 1570  &  2.70E-09 &   1920  &             &  2420  &    9.82E-10  &  2945  &    6.45E-10	 \\
   1230  & 2.94E-09 & 1580  &  2.70E-09 &   1930  &             &  2435  &    9.72E-10  &  2960  &    6.35E-10	 \\
   1240  & 4.38E-09 & 1590  &  2.76E-09 &   1940  &    1.79E-09 &  2450  &    9.59E-10  &  2975  &    6.44E-10	 \\
   1250  & 4.37E-09 & 1600  &  2.65E-09 &   1950  &    1.74E-09 &  2465  &    9.22E-10  &  2990  &    6.20E-10	 \\
   1260  & 4.04E-09 & 1610  &  2.59E-09 &   1960  &    1.69E-09 &  2480  &    9.15E-10  &  3005  &    6.12E-10	 \\
   1270  & 4.25E-09 & 1620  &  2.79E-09 &   1970  &    1.77E-09 &  2495  &    9.47E-10  &  3020  &    6.03E-10	 \\
   1280  & 4.38E-09 & 1630  &  2.62E-09 &   1980  &    1.71E-09 &  2510  &    9.58E-10  &  3035  &    6.13E-10	 \\
   1290  & 4.24E-09 & 1640  &  2.68E-09 &   2000  &    1.51E-09 &  2525  &    9.12E-10  &  3050  &    6.25E-10	 \\
   1300  & 3.08E-09 & 1650  &  2.69E-09 &   2015  &    1.49E-09 &  2540  &    9.12E-10  &  3065  &    6.24E-10	 \\
   1310  & 3.25E-09 & 1660  &  2.61E-09 &   2030  &    1.58E-09 &  2555  &    8.98E-10  &  3080  &    6.02E-10	 \\
   1320  & 4.41E-09 & 1670  &  2.60E-09 &   2045  &    1.50E-09 &  2570  &    8.85E-10  &  3095  &    6.13E-10	 \\
   1330  & 3.39E-09 & 1680  &  2.60E-09 &   2060  &    1.41E-09 &  2585  &    9.19E-10  &  3110  &    5.98E-10	 \\
   1340  & 3.85E-09 & 1690  &  2.53E-09 &   2075  &    1.40E-09 &  2600  &    8.85E-10  &  3125  &    5.86E-10	 \\
   1350  & 4.03E-09 & 1700  &  2.51E-09 &   2090  &    1.47E-09 &  2615  &    8.59E-10  &  3140  &    5.76E-10	 \\
   1360  & 3.99E-09 & 1710  &  2.28E-09 &   2105  &    1.36E-09 &  2630  &    8.53E-10  &  3155  &    5.63E-10	 \\
   1370  & 3.74E-09 & 1720  &  2.22E-09 &   2120  &    1.39E-09 &  2645  &    8.84E-10  &  3170  &    5.53E-10	 \\
   1380  & 3.77E-09 & 1730  &  2.22E-09 &   2135  &    1.36E-09 &  2660  &    8.79E-10  &  3185  &    5.30E-10	 \\
   1390  & 3.40E-09 & 1740  &  2.29E-09 &   2150  &    1.33E-09 &  2675  &    8.44E-10  &  3200  &    5.49E-10	 \\
   1400  & 3.30E-09 & 1750  &  2.29E-09 &   2165  &    1.36E-09 &  2690  &    8.55E-10  &  3215  &    5.30E-10	 \\
   1410  & 3.50E-09 & 1760  &  2.23E-09 &   2180  &    1.24E-09 &  2705  &    8.35E-10  &  3230  &    5.24E-10	 \\
   1420  & 3.50E-09 & 1770  &  2.18E-09 &   2195  &    1.26E-09 &  2720  &    8.19E-10  &  3245  &    5.02E-10	 \\
   1430  & 3.59E-09 & 1780  &  2.22E-09 &   2210  &    1.19E-09 &  2735  &    8.16E-10  &  3260  &    5.03E-10	 \\
   1440  & 3.48E-09 & 1790  &           &   2225  &    1.18E-09 &  2750  &    7.68E-10  &  3275  &    4.89E-10	 \\
   1450  & 3.55E-09 & 1800  &  2.19E-09 &   2240  &    1.23E-09 &  2765  &    7.68E-10  &  3290  &    4.79E-10	 \\
   1460  & 3.55E-09 & 1810  &  2.11E-09 &   2255  &    1.10E-09 &  2780  &    7.81E-10  &  3305  &    4.76E-10	 \\
   1470  & 3.32E-09 & 1820  &  2.09E-09 &   2270  &    1.12E-09 &  2795  &    7.32E-10  &  3320  &    4.54E-10	 \\
   1480  & 3.23E-09 & 1830  &  2.07E-09 &   2285  &    1.15E-09 &  2810  &    7.49E-10  &  3335  &    4.59E-10	 \\
   1490  & 3.21E-09 & 1840  &  1.97E-09 &   2300  &    1.14E-09 &  2825  &    7.48E-10  &  3350  &    4.84E-10	 \\
\hline
\end{tabular}
\end{table*}

\begin{table*}
\caption{HD~34816} 
\vspace{0.3cm}

\begin{tabular}{c c c c c c c c c c}
\hline
 Wavelength &  Flux &  Wavelength &  Flux &  Wavelength &  Flux &  Wavelength &  Flux &  Wavelength &  Flux   \\
\hline
   1150  &  5.79E-09  & 1500  & 2.86E-09  & 1850  &    1.63E-09  & 2315  & 9.27E-10  &    2840  &    5.18E-10 \\      
   1160  &  5.44E-09  & 1510  & 2.68E-09  & 1860  &    1.70E-09  & 2330  & 9.11E-10  &    2855  &    4.89E-10 \\   
   1170  &  4.02E-09  & 1520  & 2.64E-09  & 1870  &    1.62E-09  & 2345  & 8.79E-10  &    2870  &    4.91E-10 \\   
   1180  &  4.39E-09  & 1530  & 2.29E-09  & 1880  &    1.58E-09  & 2360  & 8.81E-10  &    2885  &    4.88E-10 \\   
   1190  &  4.45E-09  & 1540  & 2.08E-09  & 1890  &    1.46E-09  & 2375  & 8.35E-10  &    2900  &    4.67E-10 \\   
   1200  &  4.33E-09  & 1550  & 1.81E-09  & 1900  &    1.50E-09  & 2390  & 8.26E-10  &    2915  &    4.57E-10 \\   
   1210  &  2.39E-09  & 1560  & 2.17E-09  & 1910  &    1.51E-09  & 2405  & 8.18E-10  &    2930  &    4.46E-10 \\   
   1220  &  3.38E-09  & 1570  & 2.26E-09  & 1920  &              & 2420  & 8.17E-10  &    2945  &    4.44E-10 \\   
   1230  &  4.71E-09  & 1580  & 2.28E-09  & 1930  &              & 2435  & 8.05E-10  &    2960  &    4.41E-10 \\   
   1240  &  4.69E-09  & 1590  & 2.32E-09  & 1940  &    1.47E-09  & 2450  & 7.79E-10  &    2975  &    4.25E-10 \\   
   1250  &  4.59E-09  & 1600  & 2.20E-09  & 1950  &    1.39E-09  & 2465  & 7.81E-10  &    2990  &    4.29E-10 \\   
   1260  &  4.78E-09  & 1610  & 2.10E-09  & 1960  &    1.36E-09  & 2480  & 7.56E-10  &    3005  &    4.24E-10 \\   
   1270  &  4.55E-09  & 1620  & 2.23E-09  & 1970  &    1.51E-09  & 2495  & 7.39E-10  &    3020  &    4.19E-10 \\   
   1280  &  4.78E-09  & 1630  & 2.13E-09  & 1980  &    1.45E-09  & 2510  & 7.68E-10  &    3035  &    4.22E-10 \\   
   1290  &  4.87E-09  & 1640  & 2.19E-09  & 2000  &    1.32E-09  & 2525  & 7.62E-10  &    3050  &    4.31E-10 \\   
   1300  &  4.19E-09  & 1650  & 2.36E-09  & 2015  &    1.33E-09  & 2540  & 7.27E-10  &    3065  &    4.09E-10 \\   
   1310  &  4.77E-09  & 1660  & 2.22E-09  & 2030  &    1.44E-09  & 2555  & 7.09E-10  &    3080  &    4.02E-10 \\   
   1320  &  4.43E-09  & 1670  & 2.14E-09  & 2045  &    1.35E-09  & 2570  & 7.07E-10  &    3095  &    3.81E-10 \\   
   1330  &  4.12E-09  & 1680  & 2.32E-09  & 2060  &    1.24E-09  & 2585  & 6.96E-10  &    3110  &    3.89E-10 \\   
   1340  &  4.24E-09  & 1690  & 2.12E-09  & 2075  &    1.27E-09  & 2600  & 6.87E-10  &    3125  &    3.84E-10 \\   
   1350  &  4.51E-09  & 1700  & 2.15E-09  & 2090  &    1.24E-09  & 2615  & 6.48E-10  &    3140  &    3.64E-10 \\   
   1360  &  4.23E-09  & 1710  & 2.04E-09  & 2105  &    1.21E-09  & 2630  & 6.67E-10  &    3155  &    3.56E-10 \\   
   1370  &  4.34E-09  & 1720  & 1.90E-09  & 2120  &    1.23E-09  & 2645  & 6.58E-10  &    3170  &    3.59E-10 \\   
   1380  &  4.17E-09  & 1730  & 1.96E-09  & 2135  &    1.22E-09  & 2660  & 6.42E-10  &    3185  &    3.46E-10 \\   
   1390  &  3.20E-09  & 1740  & 1.97E-09  & 2150  &    1.17E-09  & 2675  & 6.14E-10  &    3200  &    3.37E-10 \\   
   1400  &  2.91E-09  & 1750  & 2.01E-09  & 2165  &    1.15E-09  & 2690  & 6.18E-10  &    3215  &    3.35E-10 \\   
   1410  &  3.52E-09  & 1760  & 2.05E-09  & 2180  &    1.10E-09  & 2705  & 6.10E-10  &    3230  &    3.24E-10 \\   
   1420  &  3.45E-09  & 1770  & 1.92E-09  & 2195  &    1.08E-09  & 2720  & 6.20E-10  &    3245  &    3.20E-10 \\   
   1430  &  3.34E-09  & 1780  & 1.98E-09  & 2210  &    1.06E-09  & 2735  & 5.88E-10  &    3260  &    3.11E-10 \\   
   1440  &  3.43E-09  & 1790  &           & 2225  &    1.07E-09  & 2750  & 5.84E-10  &    3275  &    2.94E-10 \\   
   1450  &  3.33E-09  & 1800  & 1.83E-09  & 2240  &    1.03E-09  & 2765  & 5.67E-10  &    3290  &    2.94E-10 \\   
   1460  &  3.26E-09  & 1810  & 1.78E-09  & 2255  &    1.04E-09  & 2780  & 5.60E-10  &    3305  &    2.66E-10 \\   
   1470  &  3.11E-09  & 1820  & 1.79E-09  & 2270  &    9.92E-10  & 2795  & 5.16E-10  &    3320  &    2.87E-10 \\   
   1480  &  3.00E-09  & 1830  & 1.81E-09  & 2285  &    1.01E-09  & 2810  & 5.33E-10  &    3335  &    2.86E-10 \\   
   1490  &  3.09E-09  & 1840  & 1.67E-09  & 2300  &    9.40E-10  & 2825  & 5.34E-10  &    3350  &    3.02E-10 \\   
\hline
\end{tabular}
\end{table*}

\begin{table*}
\caption{HD~214680} 
\vspace{0.3cm}

\begin{tabular}{c c c c c c c c c c}
\hline
 Wavelength &  Flux &  Wavelength &  Flux &  Wavelength &  Flux &  Wavelength &  Flux &  Wavelength &  Flux   \\
\hline
   1150  &   2.06E-09  & 1500  &   1.40E-09  & 1850  &   8.77E-10   &  2315  &   4.65E-10  & 2840  &   2.95E-10  \\
   1160  &   2.10E-09  & 1510  &   1.43E-09  & 1860  &   8.76E-10   &  2330  &   4.70E-10  & 2855  &   2.90E-10	 \\
   1170  &   1.76E-09  & 1520  &   1.40E-09  & 1870  &   8.50E-10   &  2345  &   4.36E-10  & 2870  &   2.83E-10	 \\
   1180  &   1.71E-09  & 1530  &   1.17E-09  & 1880  &   8.39E-10   &  2360  &   4.39E-10  & 2885  &   2.83E-10	 \\
   1190  &   1.73E-09  & 1540  &   1.00E-09  & 1890  &   8.34E-10   &  2375  &   4.30E-10  & 2900  &   2.68E-10	 \\
   1200  &   1.54E-09  & 1550  &   8.54E-10  & 1900  &   8.38E-10   &  2390  &   4.19E-10  & 2915  &   2.65E-10	 \\
   1210  &   7.50E-10  & 1560  &   1.10E-09  & 1910  &   8.32E-10   &  2405  &   4.29E-10  & 2930  &   2.64E-10	 \\
   1220  &   7.44E-10  & 1570  &   1.13E-09  & 1920  &   7.99E-10   &  2420  &   4.32E-10  & 2945  &   2.54E-10	 \\
   1230  &   1.75E-09  & 1580  &   1.17E-09  & 1930  &              &  2435  &   4.35E-10  & 2960  &   2.50E-10	 \\
   1240  &   1.81E-09  & 1590  &   1.16E-09  & 1940  &   7.87E-10   &  2450  &   4.26E-10  & 2975  &   2.48E-10	 \\
   1250  &   1.95E-09  & 1600  &   1.09E-09  & 1950  &   7.66E-10   &  2465  &   4.25E-10  & 2990  &   2.41E-10	 \\
   1260  &   1.98E-09  & 1610  &   9.75E-10  & 1960  &   7.26E-10   &  2480  &   4.10E-10  & 3005  &   2.43E-10	 \\
   1270  &   2.09E-09  & 1620  &   1.03E-09  & 1970  &   7.59E-10   &  2495  &   4.24E-10  & 3020  &   2.33E-10	 \\
   1280  &   2.08E-09  & 1630  &   9.68E-10  & 1980  &   7.45E-10   &  2510  &   4.08E-10  & 3035  &   2.37E-10	 \\
   1290  &   2.04E-09  & 1640  &   1.04E-09  & 2000  &   6.97E-10   &  2525  &   4.05E-10  & 3050  &   2.38E-10	 \\
   1300  &   1.92E-09  & 1650  &   1.15E-09  & 2015  &   6.75E-10   &  2540  &   4.19E-10  & 3065  &   2.28E-10	 \\
   1310  &   1.98E-09  & 1660  &   1.04E-09  & 2030  &   6.79E-10   &  2555  &   4.10E-10  & 3080  &   2.27E-10	 \\
   1320  &   1.90E-09  & 1670  &   1.03E-09  & 2045  &   6.90E-10   &  2570  &   3.91E-10  & 3095  &   2.24E-10	 \\
   1330  &   1.92E-09  & 1680  &   1.15E-09  & 2060  &   6.33E-10   &  2585  &   3.84E-10  & 3110  &   2.22E-10	 \\
   1340  &   1.83E-09  & 1690  &   1.09E-09  & 2075  &   5.85E-10   &  2600  &   3.82E-10  & 3125  &   2.16E-10	 \\
   1350  &   1.86E-09  & 1700  &   1.03E-09  & 2090  &   6.03E-10   &  2615  &   3.72E-10  & 3140  &   2.14E-10	 \\
   1360  &   1.86E-09  & 1710  &   1.04E-09  & 2105  &   5.86E-10   &  2630  &   3.66E-10  & 3155  &   2.09E-10	 \\
   1370  &   1.90E-09  & 1720  &   9.40E-10  & 2120  &   5.68E-10   &  2645  &   3.64E-10  & 3170  &   2.01E-10	 \\
   1380  &   1.87E-09  & 1730  &   9.76E-10  & 2135  &   5.51E-10   &  2660  &   3.62E-10  & 3185  &   1.95E-10	 \\
   1390  &   1.71E-09  & 1740  &   1.03E-09  & 2150  &   5.39E-10   &  2675  &   3.47E-10  & 3200  &   1.94E-10	 \\
   1400  &   1.55E-09  & 1750  &   1.02E-09  & 2165  &   5.42E-10   &  2690  &   3.44E-10  & 3215  &   1.98E-10	 \\
   1410  &   1.65E-09  & 1760  &   1.02E-09  & 2180  &   5.27E-10   &  2705  &   3.51E-10  & 3230  &   1.90E-10	 \\
   1420  &   1.65E-09  & 1770  &   9.82E-10  & 2195  &   5.03E-10   &  2720  &   3.45E-10  & 3245  &   1.85E-10	 \\
   1430  &   1.55E-09  & 1780  &   9.94E-10  & 2210  &   4.90E-10   &  2735  &   3.22E-10  & 3260  &   1.78E-10	 \\
   1440  &   1.61E-09  & 1790  &   9.62E-10  & 2225  &   5.06E-10   &  2750  &   3.36E-10  & 3275  &   1.70E-10	 \\
   1450  &   1.53E-09  & 1800  &   9.36E-10  & 2240  &   4.98E-10   &  2765  &   3.21E-10  & 3290  &   1.64E-10	 \\
   1460  &   1.49E-09  & 1810  &   8.99E-10  & 2255  &   4.75E-10   &  2780  &   3.22E-10  & 3305  &   1.70E-10	 \\
   1470  &   1.55E-09  & 1820  &   9.23E-10  & 2270  &   4.87E-10   &  2795  &   2.93E-10  & 3320  &   1.57E-10	 \\
   1480  &   1.49E-09  & 1830  &   9.54E-10  & 2285  &   4.73E-10   &  2810  &   3.07E-10  & 3335  &   1.51E-10	 \\
   1490  &   1.50E-09  & 1840  &   9.16E-10  & 2300  &   4.43E-10   &  2825  &   3.07E-10  & 3350  &   1.50E-10	 \\
\hline
\end{tabular}
\end{table*}  

%%%%%%%%%%%%%%%%%%%%%%%%%%%%%%%%%%%%%%%%%%%%%%%%%%%%%%%%%%%%%%%%%%%%%%%%%%%%%%%
\begin{figure*}
\centering
\includegraphics[width=17cm]{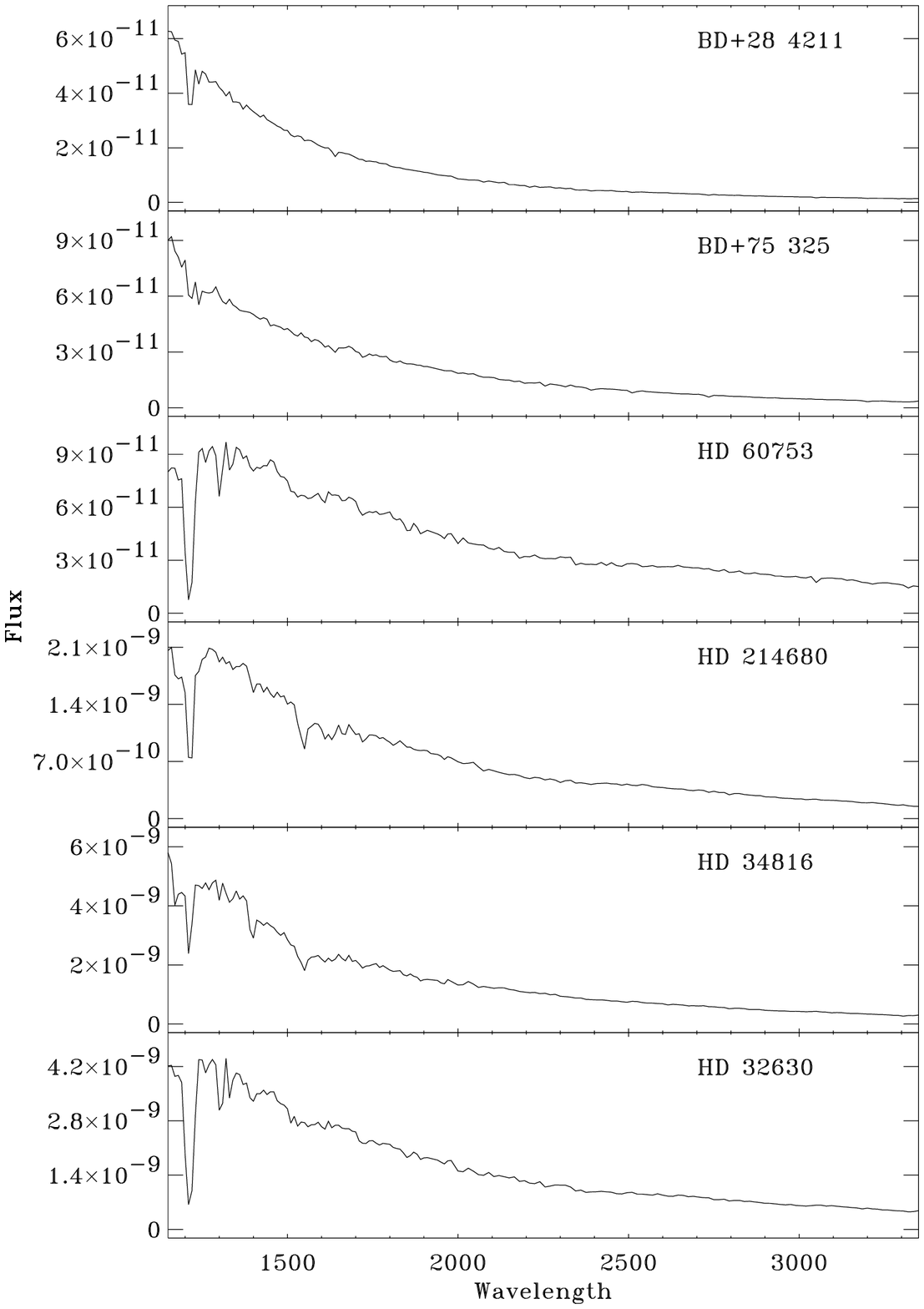}
\caption{UV spectral distribution of the IUE standard stars}
\label{fig:bd28} 
\end{figure*}
%%%%%%%%%%%%%%%%%%%%%%%%%%%%%%%%%%%%%%%%%%%%%%%%%%%%%%%%%%%%%%%%%%%%%%%%%%%%%%%

\section{The Model Fluxes of G191--B2B}

The table in Appendix B gives the model fluxes of the white dwarf G191--B2B
from Finley (private communication, 1991). These fluxes are scaled to the
OAO--2 flux scale, i.e. the fluxes provided --originally scaled to the
optical spectrophotometry of Massey et al. (1988)-- have been divided by
1.042. Wavelength is in \AA\ and flux in erg~cm$^{\rm
-2}$~s$^{\rm-1}$~\AA$^{\rm -1}$.

\begin{table*}
\caption{Absolute Fluxes of the White Dwarf G191--B2B}
\vspace{0.3cm}
 
\begin{tabular}{c c c c c c c c c c}
\hline
Wavelength &  Flux &  Wavelength &  Flux &  Wavelength	&  Flux & Wavelength	&  Flux &  Wavelength	&  Flux \\
\hline
1150   &      1.73E-11  &  1500  &   7.11E-12  &    1850   &  3.43E-12	&  2315   &  1.55E-12&	2840  & 7.38E-13 \\
1160   &      1.68E-11  &  1510  &   6.95E-12  &    1860   &  3.37E-12	&  2330   &  1.51E-12&	2855  & 7.24E-13 \\
1170   &      1.63E-11  &  1520  &   6.80E-12  &    1870   &  3.30E-12	&  2345   &  1.48E-12&	2870  & 7.11E-13 \\
1180   &      1.58E-11  &  1530  &   6.65E-12  &    1880   &  3.24E-12	&  2360   &  1.44E-12&	2885  & 6.97E-13 \\
1190   &      1.53E-11  &  1540  &   6.50E-12  &    1890   &  3.18E-12	&  2375   &  1.41E-12&	2900  & 6.84E-13 \\
1200   &      1.44E-11  &  1550  &   6.36E-12  &    1900   &  3.12E-12	&  2390   &  1.38E-12&	2915  & 6.71E-13 \\
1210   &      1.12E-11  &  1560  &   6.22E-12  &    1910   &  3.07E-12	&  2405   &  1.35E-12&	2930  & 6.59E-13 \\
1220   &      1.00E-11  &  1570  &   6.08E-12  &    1920   &  3.01E-12	&  2420   &  1.32E-12&	2945  & 6.47E-13 \\
1230   &      1.32E-11  &  1580  &   5.95E-12  &    1930   &  2.96E-12	&  2435   &  1.29E-12&	2960  & 6.35E-13 \\
1240   &      1.33E-11  &  1590  &   5.82E-12  &    1940   &  2.90E-12	&  2450   &  1.26E-12&	2975  & 6.23E-13 \\
1250   &      1.31E-11  &  1600  &   5.70E-12  &    1950   &  2.85E-12	&  2465   &  1.23E-12&	2990  & 6.12E-13 \\
1260   &      1.28E-11  &  1610  &   5.57E-12  &    1960   &  2.80E-12	&  2480   &  1.21E-12&	3005  & 6.01E-13 \\
1270   &      1.25E-11  &  1620  &   5.46E-12  &    1970   &  2.75E-12	&  2495   &  1.18E-12&	3020  & 5.90E-13 \\
1280   &      1.22E-11  &  1630  &   5.34E-12  &    1980   &  2.70E-12	&  2510   &  1.16E-12&	3035  & 5.79E-13 \\
1290   &      1.19E-11  &  1640  &   5.23E-12  &    2000   &  2.61E-12	&  2525   &  1.13E-12&	3050  & 5.69E-13 \\
1300   &      1.16E-11  &  1650  &   5.12E-12  &    2015   &  2.54E-12	&  2540   &  1.11E-12&	3065  & 5.59E-13 \\
1310   &      1.13E-11  &  1660  &   5.01E-12  &    2030   &  2.47E-12	&  2555   &  1.08E-12&	3080  & 5.49E-13 \\
1320   &      1.10E-11  &  1670  &   4.91E-12  &    2045   &  2.41E-12	&  2570   &  1.06E-12&	3095  & 5.39E-13 \\
1330   &      1.07E-11  &  1680  &   4.81E-12  &    2060   &  2.35E-12	&  2585   &  1.04E-12&	3110  & 5.30E-13 \\
1340   &      1.04E-11  &  1690  &   4.71E-12  &    2075   &  2.29E-12	&  2600   &  1.02E-12&	3125  & 5.20E-13 \\
1350   &      1.02E-11  &  1700  &   4.61E-12  &    2090   &  2.23E-12	&  2615   &  9.97E-13&	3140  & 5.11E-13 \\
1360   &      9.93E-12  &  1710  &   4.52E-12  &    2105   &  2.17E-12	&  2630   &  9.76E-13&	3155  & 5.02E-13 \\
1370   &      9.69E-12  &  1720  &   4.43E-12  &    2120   &  2.12E-12	&  2645   &  9.56E-13&	3170  & 4.94E-13 \\
1380   &      9.46E-12  &  1730  &   4.34E-12  &    2135   &  2.07E-12	&  2660   &  9.37E-13&	3185  & 4.85E-13 \\
1390   &      9.23E-12  &  1740  &   4.25E-12  &    2150   &  2.01E-12	&  2675   &  9.18E-13&	3200  & 4.77E-13 \\
1400   &      9.01E-12  &  1750  &   4.17E-12  &    2165   &  1.97E-12	&  2690   &  9.00E-13&	3215  & 4.69E-13 \\
1410   &      8.79E-12  &  1760  &   4.09E-12  &    2180   &  1.92E-12	&  2705   &  8.82E-13&	3230  & 4.61E-13 \\
1420   &      8.58E-12  &  1770  &   4.01E-12  &    2195   &  1.87E-12	&  2720   &  8.64E-13&	3245  & 4.53E-13 \\
1430   &      8.38E-12  &  1780  &   3.93E-12  &    2210   &  1.83E-12	&  2735   &  8.47E-13&	3260  & 4.45E-13 \\
1440   &      8.18E-12  &  1790  &   3.85E-12  &    2225   &  1.78E-12	&  2750   &  8.30E-13&	3275  & 4.38E-13 \\
1450   &      7.99E-12  &  1800  &   3.78E-12  &    2240   &  1.74E-12	&  2765   &  8.14E-13&	3290  & 4.31E-13 \\
1460   &      7.81E-12  &  1810  &   3.71E-12  &    2255   &  1.70E-12	&  2780   &  7.98E-13&	3305  & 4.23E-13 \\
1470   &      7.62E-12  &  1820  &   3.63E-12  &    2270   &  1.66E-12	&  2795   &  7.83E-13&	3320  & 4.16E-13 \\
1480   &      7.45E-12  &  1830  &   3.57E-12  &    2285   &  1.62E-12	&  2810   &  7.67E-13&	3335  & 4.10E-13 \\
1490   &      7.28E-12  &  1840  &   3.50E-12  &    2300   &  1.58E-12	&  2825   &  7.53E-13&	3350  & 4.03E-13 \\
\hline  		   
\end{tabular}
\end{table*}

%%%%%%%%%%%%%%%%%%%%%%%%%%%%%%%%%%%%%%%%%%%%%%%%%%%%%%%%%%%%%%%%%%%%%%%%%%%%%%%
\begin{figure*}
\includegraphics[width=17cm]{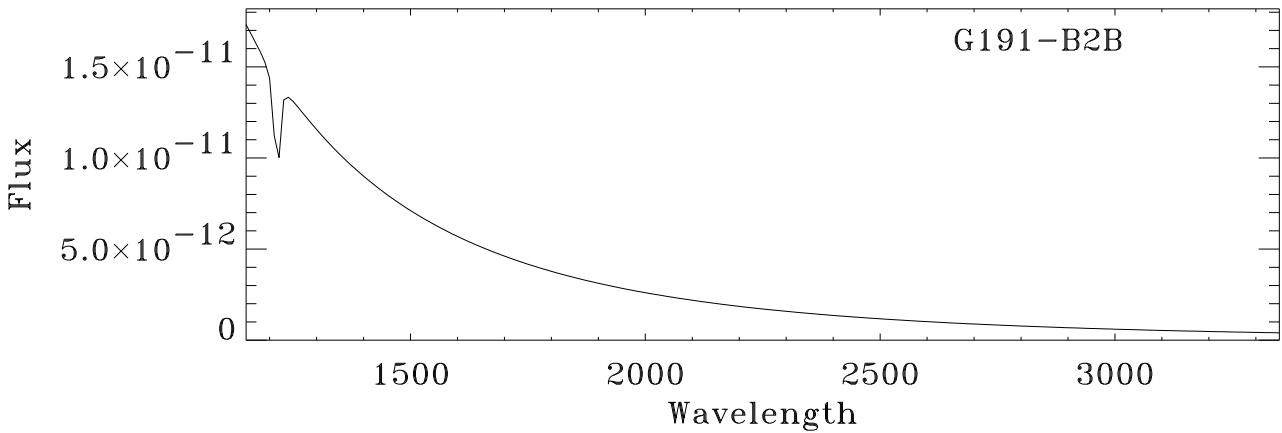} 
\caption{UV spectral distribution of the White Dwarf G191--B2B}
\label{fig:g191} 
\end{figure*}
%%%%%%%%%%%%%%%%%%%%%%%%%%%%%%%%%%%%%%%%%%%%%%%%%%%%%%%%%%%%%%%%%%%%%%%%%%%%%%%


\begin{thebibliography}{}

\bibitem[{Barstow et al.}{1993}]{barstow93}
Barstow, M.A., Fleming, T.A., Finley, D.S., Koester, D.,
Diamond, C.J., 1993, MNRAS 260, 631

\bibitem[{Beeckmans, F.}{1977}]{beeckmans}
Beeckmans, F., 1977, A\&A 60,1

\bibitem[{Bohlin and Holm}{1984}]{bholm}
Bohlin, R.C., Holm, A.V., 1984, ESA-IUE Newsletter 20, 22

\bibitem[{Bohlin et al.}{1980}]{bohlin80}
Bohlin, R.C, Holm, A.V., Savage, B.D., Snijders, M.A.J., Sparks,
W.M., 1980, A\&A 85,1

\bibitem[{Bohlin et al.}{1995}]{bohlin95}
Bohlin, R.C., Colina, L., Finley, D.S., 1995, AJ 110, 1316

\bibitem[{Bohlin}{1996}]{bohlin96}
Bohlin, R.C., 1996, AJ 111, 1743. 

\bibitem[{Bohlin}{2000}]{bohlin00}
Bohlin, R.C., 2000, AJ 120, 437. 

\bibitem[{Bruhweiler}{1991}]{bru81}
Bruhweiler, F.C., 1991, Report to the IUE Three Agency Meeting, June 1991

\bibitem[{Bruhweiler and Kondo}{1981}]{brukon81}
Bruhweiler, F.C., Kondo, Y., 1981, ApJ 248, L123

\bibitem[{Brune et al.}{1979}]{brune}
Brune, W. H., Mount, G. H.,  Feldman, P. D., 1979, ApJ 227, 884

\bibitem[{Cassatella 1990}{1990}]{cass90}
Cassatella, A., 1990, Report to the IUE Three Agency Meeting, May 1990

\bibitem[{Cassatella et al.}{2000}]{cass00}
Cassatella, A., Altamore, A., Gonz\'alez-Riestra, R., 
et al., 2000, A\&AS 141, 331

\bibitem[{Colina and Bohlin}{1994}]{colina94}
Colina, L., Bohlin, R.C., 1994, AJ 108, 1931

\bibitem[{Finley}{1993}]{finley93}
Finley, D.S., 1993, in ``Calibrating HST'', ed. J.C. Blades \&\
S.J. Osmer, p. 416

\bibitem[{Finley at al.}{1990}]{finley90}
Finley, D. S., Basri, G.,  Bowyer, S., 1990, ApJ 359, 483

\bibitem[{Finley at al.}{1997}]{finley97}
Finley, D.S., Koester,D., Basri,G., 1997, ApJ 488, 375

\bibitem[{Garhart}{1991}]{thda}
Garhart, M.P., 1991, Report to the IUE Three Agency Meeting,
June 1991, p. VI-38

\bibitem[{Garhart et al.}{1997}]{garhart97}
Garhart, M.P., Smith, M.A., Levay, K.L., Thompsom, R.W., 1997,
``International Ultraviolet Explorer New Spectral Image
Processing Information Manual, Version 2.0''

\bibitem[{Gonz\'alez-Riestra}{1991}]{crt}
Gonz\'alez-Riestra, R., 1991, Report to the IUE Three Agency
Meeting, June 1991, p. VI-82

\bibitem[{Gonz\'alez-Riestra}{1998}]{nonl}
Gonz\'alez-Riestra, R., 1998, in ``Ultraviolet Astrophysics beyond the
IUE Final Archive'', eds. W. Wamsteker and R. Gonz\'alez--Riestra, ESA
SP-413, p. 703 

\bibitem[{Gonz\'alez-Riestra et al.}{2000}]{high}
Gonz\'alez-Riestra, R., Cassatella, A.,  Solano, E.,  Altamore, A.,
Wamsteker, W., 2000, A\&AS 141, 343

\bibitem[{Greenstein and Oke}{1979}]{greenstein}
Greenstein, J. L., Oke, J. B., 1979, ApJ 229, L141

\bibitem[{Hackney et al.}{1982}]{hackney}
Hackney, R. L., Hackney, K. R. H.,  Kondo, Y., 1982, in ``Advances in
Ultraviolet Astronomy'', NASA CP-2238,  335

\bibitem[{Kimble et al.}{1993}]{nh}
Kimble, R.A., Davidsen, A.F., Blair, W.P., et al., 1993, ApJ 404,
663

\bibitem[{Kruk et al.}{1997}]{kruk97}
Kruk, J.W., Kimble, R.A., Buss, R.H., et al., 1997, ApJ 482, 546

\bibitem[{Kruk et al.}{1999}]{kruk99}
Kruk, J.W., Brown, T. M., Davidsen, A.F., et al., 1999, ApJS 122,
299

\bibitem[{Massey et al.}{1988}]{massey88}
Massey, P., Strobel, R., Barnes, J.V., Anderson, E., 1998, ApJ
328, 315

\bibitem[{Meade}{1978}]{meade78}
Meade, M., 1978, private communication.

\bibitem[{Oliversen}{1987}]{cdct}
Oliversen, N., 1987, Report to the IUE Three Agency Meeting, November 1987

\bibitem[{de la Pena}{1992}]{1515}
de la Pe\~na, M., 1992, Report to the IUE Three Agency Meeting, June 1992

\bibitem[{P\'erez--Calpena and Pepoy}{1997}]{finalreport}
P\'erez--Calpena, A., Pepoy, J., 1997, ``IUE Spacecraft Operations: Final
report'', ESA SP-1215

\bibitem[{Rodriguez-Pascual et al.}{1999}]{pmr}
Rodr\'{\i}guez-Pascual, P.M., Gonz\'alez-Riestra, R., Schartel,
N., Wamsteker, W.,  1999, A\&AS 139, 183

\bibitem[{Schartel and Skillen}{1998}]{scsk}
Schartel, N., Skillen, I., 1998, in ``Ultraviolet Astrophysics beyond the
IUE Final Archive'', eds. W. Wamsteker and R. Gonz\'alez--Riestra, ESA
SP-413, p. 735 

\bibitem[{Vennes}{1992}]{vennes}
Vennes, S., 1992, ApJ 390, 590

\bibitem[{Wamsteker}{2000}]{news}
Wamsteker, W., 2000, ``The INES Newsleter'', ESA

\bibitem[{Wamsteker et al.}{2000}]{ines}
Wamsteker, W.,  Skillen, I., Ponz, J.D., et al., 2000, 
A\&SS, 273, 155
 
\bibitem[{Wolff et al.}{1998}]{abun}
Wolff, B., Koester, D., Dreizler, S., Hass, S., 1998, A\&A 329,
1045

\end{thebibliography}
\end{document}